\documentclass{sig-alternate-2013}
\newfont{\mycrnotice}{ptmr8t at 7pt}
\newfont{\myconfname}{ptmri8t at 7pt}
\let\crnotice\mycrnotice%
\let\confname\myconfname%

\permission{} 
\conferenceinfo{University of Helsinki, Department of Computer Science, }{Aug.2015.} 
\copyrightetc{This work is accepted for publication in ACM MsWiM 2015}

\usepackage{multirow}
\usepackage{algorithm}
\usepackage{algorithmic}

\usepackage{amsmath}

\usepackage{amssymb}
\usepackage{graphicx}
\usepackage{marvosym}
\usepackage{multirow}
\usepackage{tabularx}
\usepackage[table]{xcolor}
\usepackage{subfigure}
\usepackage{url}
\usepackage[titletoc,title]{appendix}
\usepackage{cite}
\usepackage{cases}
\usepackage{setspace}
\usepackage{pifont}% 

\usepackage{enumitem}
\setitemize{noitemsep,topsep=0pt,parsep=0pt,partopsep=0pt}
\setlist[itemize]{leftmargin=*}

\newcommand{\avail}{\alpha}
\newcommand{\setNdis}{\mathcal{N}}
\newcommand{\Ndis}{N}

\newcommand{\setNtagged}{\mathcal{M}}
\newcommand{\Nodes}{N}
\newcommand{\setNodes}{\mathcal{N}}

\usepackage{enumitem}

\makeatletter
 \def\@textbottom{\vskip \z@ \@plus 1pt}
 \let\@texttop\relax
\makeatother

\newcommand{\leftq}{``}
\begin{document}

\title{Two Hops or More: \\On Hop-Limited Search in Opportunistic Networks}
\numberofauthors{4}
\auwidth38mm  %
\author{
\alignauthor
Suzan Bayhan\\
       \affaddr{University of Helsinki}\\
       \affaddr{Finland}\\
       \email{bayhan@hiit.fi}
\alignauthor 
Esa Hyyti\"a\\
       \affaddr{Aalto University}\\
       \affaddr{Finland}\\
       \email{esa@netlab.tkk.fi}
\alignauthor
Jussi Kangasharju\\
       \affaddr{University of Helsinki}\\
       \affaddr{Finland}\\
       \email{jakangas@cs.helsinki.fi}
\alignauthor 
J\"org Ott\\
       \affaddr{Aalto University}\\
       \affaddr{Finland}\\
       \email{jo@netlab.tkk.fi}
}

\maketitle

\maketitle
\begin{abstract}
While there is a drastic shift from host-centric networking to content-centric networking, 
how to locate and retrieve the relevant content efficiently, especially in a mobile network, is still an open question. Mobile devices host increasing volume of data which could be shared with the nearby nodes in a multi-hop fashion. However, searching for content in this resource-restricted setting is not trivial due to the lack of a content index, as well as, desire for keeping the search cost low. In this paper, we analyze a lightweight search scheme, \textit{hop-limited search}, that forwards the search messages only till a maximum number of hops, and requires no prior knowledge about the network. We highlight the effect of the hop
limit on both search performance (i.e., success ratio and delay) and
associated cost along with the interplay between content
availability, tolerated waiting time, network density, and mobility.  Our analysis, using the real mobility traces, as well as synthetic models, shows
that the most substantial benefit is achieved at the first few hops
and that after several hops the extra gain diminishes as a function of content availability and tolerated delay. We also observe that the
\textit{return path} taken by a response is on average longer than the
\textit{forward path} of the query and that the search cost 
increases only marginally after several hops due to the small network diameter.
 
\end{abstract}
 
\section{Introduction}
 
Mobile devices can establish opportunistic networks---a flavour of
Delay-tolerant Networks (DTNs) \cite{fall2003delay}---among each other
in a self-organising manner and facilitate communication without the
need for network infrastructure.  The capabilities of today's smart
mobile devices yield substantial computing and storage resources and
means for creating, viewing, archiving, manipulating, and sharing
content. In addition, content downloaded from Internet is recommended to be cached at the handset~\cite{qian2012web}. This yields a huge reserve of data stored in mobile devices,
which users may be willing to share.  While this
is typically done using Internet-based services, opportunistic networks
enable content sharing among users in close proximity of each other.
 
In this paper, we focus on a human-centric DTN, where nodes search for
information stored in some of the nodes.  
The nodes lack a global view
of the network, i.e., there is no service that could index the stored
content and assist a searching node in finding content (or indicate
that the sought information does not exist).
This means that operations are decentralized and, as mobile nodes have
resource constraints (e.g., energy), we need to control the spread of
the messages when searching.
One such control mechanism is imposing the maximum number of hops a
message can travel. 
We call a search scheme that limits the message's
path to maximum $h$ hops as $h$-hop search. 
\textit{Hop-limited search}~\cite{vojnovic2011hop,hyytia2014searching} is of our interest as it is a lightweight scheme that does not require intensive information collection about the nodes or the content items. 
However, determining the optimal $h$ is not straightforward. 
Although a large $h$ tends to increase replication, it also increases the chance of finding the desirable target (content or searching node), thereby decreasing replication. 

While many works in the literature apply hop limitations~\cite{vojnovic2011hop,boldrini2012less}, %
mostly two-hop such as \cite{al2008performance}, to the designed routing protocols, the motivation behind setting a particular hop value is not clear. 
In \cite{ahops_icc2015}, we analyze the effect of hop limit on the routing performance by modelling the optimal hop limited routing as \textit{all hops optimal path} problem~\cite{guerin2002computing}.
However, opportunistic search necessitates considering the availability of the sought content as well as routing of the response to the searching node. 
In our previous work~\cite{hyytia2014searching}, we analytically modelled the search utility and derived the optimal hop count for a linear network, e.g., search flows through a single path and response messages follow the same path.
In this paper, we provide an elaborate
analysis of hop-limited search in a mobile opportunistic
network considering the search success, completion time, and the cost.

Search is a two-phase process. We refer to the first phase in which query is routed towards the content providers as \textit{forward path} and the second phase in which response is routed towards the searching node as \textit{return path}. First, we present an analysis on the forward path via an analytical model. We show the interplay between tolerated waiting time (how long the searching node can wait for the forward path), content availability, 
and the hop count providing the maximum search success ratio for a specific setting. Next, we verify our analysis of the forward path and elaborate also on the return path via simulations.

Our results suggest the following:
\begin{itemize} 
\item Generally speaking, search performance increases with increasing $h$ especially for scarcely available content. However, the highest improvement is often achieved at the second hop. After that, the improvements become smaller and practically diminish when $h>5$.
\item We observe that return path requires on average longer hops/time compared to the forward path, which may imply that optimal settings for the forward path does not yield the optimal performance on the return path. 
\item We show that the search cost first increases and after several hops ($h\approx 4$) it tends to stabilise. This is due to the small diameter of the network; even if $h$ is large letting the nodes replicate a message to other nodes, each node gets informed about the search status quickly so that replication of obsolete messages is stopped.
\end{itemize}

The rest of the paper is organised as follows. Section~\ref{sec:SysModel} introduces the considered system model followed by Section~\ref{sec:HopLimited} introducing the hop-limited search. Section~\ref{sec:NumericalAnalysis} numerically analyzes the forward path, while
 Section~\ref{sec:Sims} focuses on the whole search process. Section~\ref{sec:Related} overviews the related work and highlights the points that distinguish our work from the others. Finally, Section~\ref{sec:Conclusions} concludes the paper.

\section{System Model}\label{sec:SysModel}

\begin{figure}
\centering
\includegraphics[scale=0.37]{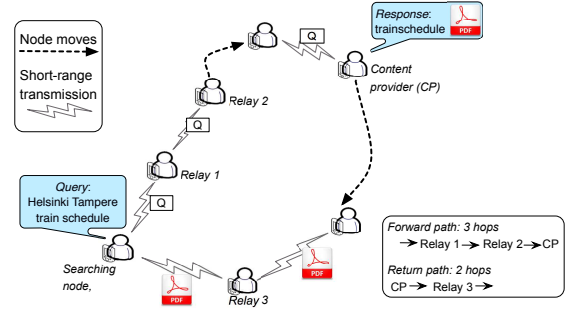} \caption{Network model, forward and return paths.}\label{fig:model} 
\end{figure}

We consider a mobile opportunistic network of $\Nodes$ nodes as in Fig.~\ref{fig:model}. Nodes move according to a mobility model which results in i.i.d.\ meeting rates %
 between any two nodes. We use the following terminology: %

\textbf{Searching node ($n_s$)} is a node that initiates the search for a content item.  %
We assume that content items are equally likely to be sought, i.e., uniform \textit{content popularity}. In
    reality, content items have diverse popularities; e.g., YouTube
    video popularity follows a Zipf distribution~\cite{gill2007youtube}.
    We choose uniform distribution to avoid a bias toward the ``easy''
    searches.

\textbf{Tagged node} is a node that holds a copy of the sought content. Only a fraction $\alpha$ of the nodes are tagged, where $\alpha$ is referred to as the \textit{content availability}. Although it is expected that search dynamics such as caching changes $\avail$, we assume that it does not change over time. The sets of all 
and tagged nodes are denoted by $\setNodes$ and $\setNtagged$, and their sizes $N$ and $M$, respectively. 
The number of tagged nodes is $M=\avail\Nodes$. 
Every node is equally likely to be a tagged node.

\textbf{Forward path and return path:} In the first step of the search, a query is disseminated in the network to find a tagged node. In case the first step is completed, a response generated by a content provider is routed back towards $n_s$ in the second step. We refer to the former as the \textit{forward} or \textit{query path} and to the latter as the \textit{return} or \textit{response path}.

\textbf{Tolerated waiting time ($T$)} is the maximum duration to find the content~(excluding the return path). 
Note that total tolerated waiting time is $2T$ if we assume the same delay restriction for the return path.

\section{Hop-Limited Search}\label{sec:HopLimited}
To describe the basic operation of hop-limited search, assume that $n_s$ creates a query that includes information about its search at time $t=0$. When $n_s$ encounters another node that does not have this query, $n_s$ replicates the query to it to reach a tagged node faster. A node acquiring a copy of the message becomes a \textit{discovered node}. The discovered node also starts to search and replicate the query to \textit{undiscovered nodes}. Each message contains a header, referred to as \textit{hop count}, representing %
 the number of nodes that forwarded this message so far. Messages at $n_s$ are 0-hop messages; those received directly from $n_s$ are $1$-hop messages. 
 We refer to a search scheme as $h$-hop search if a search message can travel at most $h$ hops. Hence, when a node has $h$-hop message, it does not forward it further. We also call a node with a $i$-hop message as the $i$-hop node for that particular message. In $h$-hop search, when an $i$-hop and $j$-hop node meet (without loss of generality we assume $i<j\leq h$), the state of $j$-hop node is updated to $(i+1)$-hop if $j>i+1$. The forward path completes when a copy of the query reaches a tagged node.

\subsection{Search Success Ratio}\label{sec:ForwardPathSuccessRatio}
Let us first consider a search scheme that does not put any hop limitation but instead limits the total number of replications on the forward path 
and the return path to $K$ and $K'$, respectively. 
For a content item with availability $\avail$, we can calculate the forward success ratio of this search scheme as:
\begin{align}
\textrm{Forward success ratio} &= 1-(1-\alpha)^K \label{eq:Pfw}.
\end{align}
Similarly, we calculate the search success ratio, i.e., both steps are completed, as:
\begin{align}
P_{s}= \sum_{k=1}^{K}&Pr\{\textrm{k content providers are discovered}\} \nonumber\\
&\times Pr\{\textrm{at least one of k responses reaches } n_s\} \nonumber.
\end{align}
We can expand the above formulation which leads to:
\begin{align}
P_{s}{=}\sum_{k=1}^{K} \binom{K}{k}\avail^k (1{-}\avail)^{K-k} \left(1{-}(1-\frac{K'}{N-1})^k\right).\label{eq:Ps}
\end{align}
Let $\gamma=\frac{K'}{N-1}$, i.e., the probability that a response reaches $n_s$. After replacing $\gamma$ into (\ref{eq:Ps}), we apply some manipulations by the help of binomial theorem: that is 
$(x+y)^n = \sum_{k=0}^{n}x^ky^{n-k}.$ 
Then, we find the search success as:
\begin{align}
P_{s}= 1- (1-\avail\gamma)^{K} \label{eq:Ps_simplified}.
\end{align}
Please refer to Appendix~\ref{appendix} for the details of the above derivation.

\begin{figure} 
\centering
\includegraphics[width=0.39\textwidth]{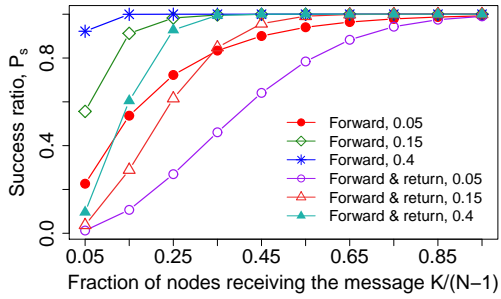}
\caption{Search success with increasing degree of replication $K$ for $\avail=\{0.05, 0.15, 0.40\}$.}
\label{fig:search_success_M}
\end{figure}

Fig.~\ref{fig:search_success_M} plots the success ratio with increasing fraction of nodes that receive the message. We plot the success of both the forward path and the total search under various content availability values. The results are for equal number of replications for the forward and return path, i.e., $K'=K$. 
The figure shows that to ensure a desirable level of success, search has to cover certain fraction of nodes, which depends on the content availability. 
In hop-limited search, there is no explicit restriction on number of replications $K$, but rather it is implicitly set by $h$ and $T$. 
This result brings us to the question of how search coverage in the number of nodes changes with $h$ and $T$.

Let $\setNdis_{h}(t)$ be the set of discovered nodes excluding $n_s$ at time $t$ and $\Ndis_{h}(t)=|\setNdis_{h}(t)|$ its size. Faloutsos \textit{et al.}~\cite{faloutsos1999power} define $\Ndis_{h}$ for a static graph as node neighborhood within $h$ hops.
Similarly, we define $\Ndis_{h}(T)$ as the number of nodes that can be reached from a source node less than or equal to $h$ hops under time limitation $T$. 
Let $P_h(T)$ denote the \textit{forward path success ratio} defined as the probability that a query reaches one of the tagged nodes in a given time period $T$ under $h$-hop limitation.
For a search that seeks a content item with availability $\alpha$, we approximate $P_{h}(T)$ as follows:
\begin{align}
P_h(T) &= \sum_{n=1}^{N}P\{\Ndis_{h}(T)=n\}\cdot(1-(1-\alpha)^n) \nonumber\\
&=1-E[(1-\alpha)^{\Ndis_{h}(T)}] \nonumber\\
&\approx 1-(1-\alpha)^{E[\Ndis_{h}(T)]} \label{eq:Pcompletion}.
\end{align}
Note that (\ref{eq:Pcompletion}) provides an upper bound for $P_h(T)$ and will only be used to understand the effect of $h$ and $T$.
In numerical evaluations, we relax all the simplifications (e.g., i.i.d meeting rates) and 
experiment using real mobility traces.
Given (\ref{eq:Pcompletion}), our problem reduces to discovering how $E[\Ndis_{h}(T)]$ changes with $h$ and $T$. 
However, as observed in \cite{Gao_FwRedundancy_Infocom2014},
each meeting may not lead to a new node being discovered, i.e., some nodes may meet again or a node may be met by several nodes. 
Therefore, total meetings in time period $T$ results in a very optimistic estimate for $E[\Ndis_{h}(T)]$. 
In addition to the overlaps of opportunistic contacts, the hop restriction may result in lower $E[\Ndis_{h}(T)]$.
 In contrast to static networks~\cite{LadaLocalSearch}, modelling $E[\Ndis_{h}(T)]$ for general time-evolving networks is not straightforward. Therefore, we derive $E[\Ndis_{h}(T)]$ from mobility traces and plug it into (\ref{eq:Pcompletion}). 
 See also \cite{hyytia_critic_infocom2013} for an analysis of how mobility and node density affect the communication capacity of a mobile opportunistic network.

\subsection{Benefit of One Additional Hop}
We define the effect of increasing hop count from $h$ to $h+1$ on the search performance as the \textit{added benefit} and calculate it as the difference between $P_h(T)$ and $P_{h+1}(T)$: $\Delta P_{h,h+1} = P_{h+1}(T) - P_{h}(T)$, where $P_{h}(T)$ values are either approximated as above or obtained from experiments.  
Given that users can perceive only significant improvements rather than minimal changes in performance, we are more interested in larger values $\Delta P_{h,h+1}$.
Let $h_{\beta}$ denote the hop count beyond which the added benefit of one additional hop is lower than $\beta$. 
More formally, $h_{\beta}$ is defined as: 
\begin{align}
h_{\beta} \triangleq \arg\max_h\left(\Delta P_{h,h+1}\geq \beta\right) + 1.
\end{align}
While $\beta\to 0$ yields the optimal hop count $h^{*}$ achieving the highest performance, i.e., $h_{\beta}=h^{*}$, we search for $h$ values that lead to higher $\beta$. 
Setting $\beta\to 0$, we assess until which hop there is still some gain, albeit minimal, whereas higher $\beta$ values help discovering the last hop with substantial benefit. We refer to these $\beta$ values as \emph{any-benefit} and \emph{fair-benefit} hop, respectively.

\subsection{Time to Forward Path Completion}\label{sec:TTsearchcompletion}

In this section, we derive the \textit{forward path completion time} which is the time required for an initiated query to reach one of the content providers. 
Let us denote by $m_i(t)$ the total number of $i$-hop nodes at time $t$. These nodes have received the message via ($i-1$) relays. Subsequently, we denote the state of a Continuous Time Markov Chain (CTMC) by:
$$S(t)=(m,m_0(t),m_1(t),\ldots,m_i(t),\ldots,m_{h-1}(t),m_{h}(t))$$
where
\begin{itemize}
\item $m=\Ndis_h(t)+1$ is the number of nodes with a copy of the query ($n_s$ and all discovered nodes) and
\item $m_i(t)$ nodes are $i$-hop nodes and $\sum_{i=0}^{h} m_i(t)=m$. 
\end{itemize}

States that have $m$ nodes holding the search query are represented as $S_m$.
 The state space consists of all $S_{m}$ transient states where $m\in\{1,\ldots,N-M\}$ and the absorbing state $S_{tagged}$, which represents discovery of a tagged node.
Hereafter, we drop the time parameter for
clarity. From state $S_m$, three types of events trigger state transitions:
\begin{itemize}
\item\textit{Meeting a tagged node}: An arbitrary $i$-hop node in $\setNdis_h$ where $i<h$ meets with one of the tagged nodes in $\setNtagged$. The resulting state is  $S_{tagged}$ and the search ends. There is only one such transition from every state.
\item \textit{Meeting a node that is undiscovered and not tagged}: An $i$-hop node meets an undiscovered untagged node. There are $\sum_{i=0}^{h-1}1_{[m_i>0]}$ such transitions from this state where indicator function $1_{[f(\cdot)]}$ yields 1 if $f(\cdot)$ evaluates to \textit{true}, and the resulting state is  $S_{m+1}=(m+1,\ldots,m_{i+1}+1,\ldots)$. 
\item \textit{Meeting among discovered nodes}: We call a meeting between $i$-hop and $j$-hop node an \textit{$(i,j)$-meeting} where $i<j$.
If $j>i+1$, $\Ndis_{h}$ does not change but $m_j$ and $m_{i+1}$ change. The new state is $S_m^{'}=(m,\ldots,m_{i+1}+1,\ldots, m_j-1,\ldots)$. The number of such transitions from a state $S_m$ is
$\sum_{i=0}^{h-2}1_{[m_i>0]}\left(\sum_{j=i+2}^{h}1_{[m_j>0]}\right).$
\end{itemize}
We call these three events \textit{Type-tagged}, \textit{Type-1}, and \textit{Type-0} events; the corresponding transition rates are denoted by $\lambda_{\mathrm{tagged}}$, $\lambda_{1}$, and $\lambda_{0}$, respectively. If the Type-1 event is a meeting with an $i$-hop node, we denote the respective rate as $\lambda_{1}^{i}$. Similarly, if a Type-0 event is due to an $(i,j)$-meeting, the rate is $\lambda_0^{i,j}$. For state $S_m=(m,m_0,m_1,\ldots,m_i,\ldots,m_{h})$ the transitions leading to a state change are:
\begin{align}
S_m &\xrightarrow{\lambda_{\mathrm{tagged}}} S_{\mathrm{tagged}} \label{eq:TransitionTagged} \\
S_m &\xrightarrow{\lambda_{1}^{i}} (m+1,\ldots,m_{i+1}+1,\ldots,m_h)  \label{eq:Transitioni}  \\
S_m &\xrightarrow{\lambda_{0}^{i,j}} (\ldots,m_{i+1}+1,\ldots, m_j-1,\ldots,m_h)   \label{eq:Transitionij}
\end{align} and the corresponding transition rates are:
\begin{align}
\lambda_{\mathrm{tagged}} &= \lambda (m-m_h)M  \label{eq:LambdaTagged} \\
\lambda_{1}^{i} &=  \lambda m_i (N-M-m)  \label{eq:Lambdai}\\
\lambda_{0}^{i,j} &= \lambda  m_i m_j \label{eq:Lambdaij} 
\end{align}
where $\lambda$ is the pairwise meeting rate.

Since solving the given Markov model may not be practical due to the state space explosion for large $h$ and $N$, we approximate $T_h$ and show its high accuracy in Section~\ref{sec:NumericalAnalysis} by comparing it with the results of Markov model. 
Let $T(m,h)$ denote the remaining time to search completion under $h$-hop search and when $m$ nodes hold the search query. Under $h$-hop search, only $m_{h^{-}}= m-m_h$ nodes are actively searching and can forward the query to their encounters. Assume that $m_i$ are identical for all $i\geqslant 1$. %
Then, the number of searching nodes $m_{h^{-}}$ is:
\begin{align}
m_{h^{-}} &= 1 + (m-1)(1-\frac{1}{h}) \nonumber.
\end{align}
We calculate $T(m,h)$ as:
\begin{align}
T(m,h) = \frac{1}{\lambda m_{h^{-}}} + \frac{\alpha}{\lambda}0 + \frac{\lambda-\alpha}{\lambda}T(m+1,h) .
 \label{eq:Tmh}
\end{align}
We expand $T(m+1,h)$ similarly and substitute in (\ref{eq:Tmh}). After a certain number of nodes are searching, the remaining time to search completion converges to zero, i.e., $T(m+k+1,h)\to 0$. Denote $q=(1-\alpha/\lambda)$. Then, we find:
\begin{align}
T(m,h) = \sum_{i=0}^{\infty}\frac{q^{i}}{\lambda (1+(m+i-1)(1-{h}^{-1}))}
 \nonumber %
\end{align}
Solving for $m=1$ gives an approximation for $T_h$: %
\begin{align}
\tilde{T}_h = \sum_{i=0}^{\infty}\frac{q^{i}}{\lambda (1+i(1-{h^{-1}}))} .
 \label{eq:TmhApproxm1} 
\end{align}

\section{Forward Path}\label{sec:NumericalAnalysis}

In this section, we evaluate the performance of hop-limited search~(referred to as HOP) while varying (i) content availability, (ii) tolerated waiting time, (iii) network density, and (iv) mobility scenarios. We use $\alpha=\{0.40, 0.15, 0.05\}$ for \textit{high}, \textit{medium}, and \textit{low} content availability, respectively.

\subsection{Datasets}
In our analysis, we use real traces of both humans and vehicles. For the former, we use the Infocom06 dataset~\cite{cambridge-haggle-2006-01-31} which represents the traces of human contacts during Infocom 2006 conference. For the latter, we use the Cabspotting dataset~\cite{epfl-mobility-2009-02-24} that stores the GPS records of the cabs in San Francisco. To gain insights about more general network settings, we also analyze a synthetic mobility model that reflects realistic movement patterns in an urban scenario. %
 Below, we overview the basic properties of each trace:

\textbf{Infocom06:} This data set records opportunistic Bluetooth contacts of 78 conference participants who were carrying iMotes and 20 static iMotes for the duration of the conference, i.e., approximately four days. In our analysis, we treated all devices as identical nodes which host content items and initiates search queries. This trace represents a small network in which people move in a closed region. %

\textbf{Cabspotting:} This data set records the latitude and longitude information of a cab as well as its occupancy state and time stamp. The trace consists of updates of the 536 cabs moving in a large city area  for a duration of 30 days. For our analysis, we focused only on the first three days and a small region of approximately 10\,km$\times$10\,km area. 496 cabs appear in this region during the specific time period. The cab information is not updated at regular intervals. Hence, we interpolated the GPS data so as to have an update at every 10 s for each cab. Next, we set transmission range to 40 m to generate contacts among cabs. This trace represents an urban mobility scenario~\cite{epfl-mobility-2009-02-24, hoque2014efficient}.

\textbf{Helsinki City Scenario (HCS):} This setting represents an opportunistic human network in which the walking speed is uniformly distributed in [0.5,1.5]\,m/s. The nodes move in a closed area of 4.5\,km$\times$3.4\,km. HCS~\cite{keranen-theone} uses the downtown Helsinki map populated with \textit{points-of-interests} 
(e.g., shops), between which pedestrians move along shortest path.

\subsection{Forward Path Success Ratio}
We derive the neighborhood size $\Ndis_{h}(T)$ as an average of 500 samples where each sample represents an independent observation of the network starting at some random time, from an arbitrary node and spanning an observation window equal to the tolerated waiting time. Next, we calculate  $P_h(T)$ using (\ref{eq:Pcompletion}).
 We use R~\cite{R} for these analysis. In the following, we mostly focus on the more challenging cases such as short $T$ or low $\avail$, and report the representative results due to the space limitations.

\begin{figure}
       \centering
      \subfigure[Infocom06, $\Ndis_{h}(T)$.]{
         \includegraphics[scale=0.22]{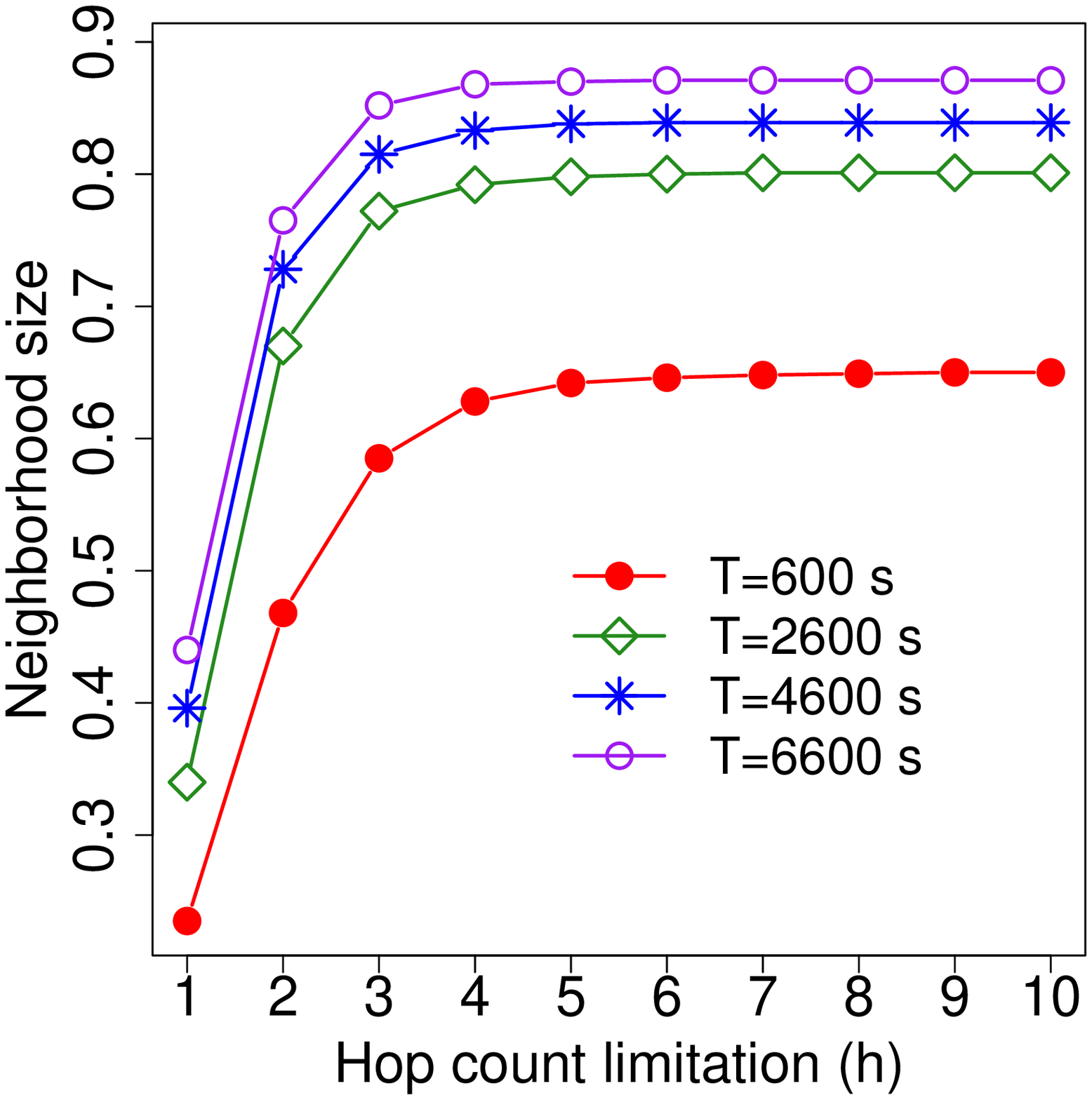}
         \label{fig:Infocom06NodeUnionByHop_98}}
       \subfigure[Infocom06, $P_h(T)$.]{
                \includegraphics[scale=0.22]{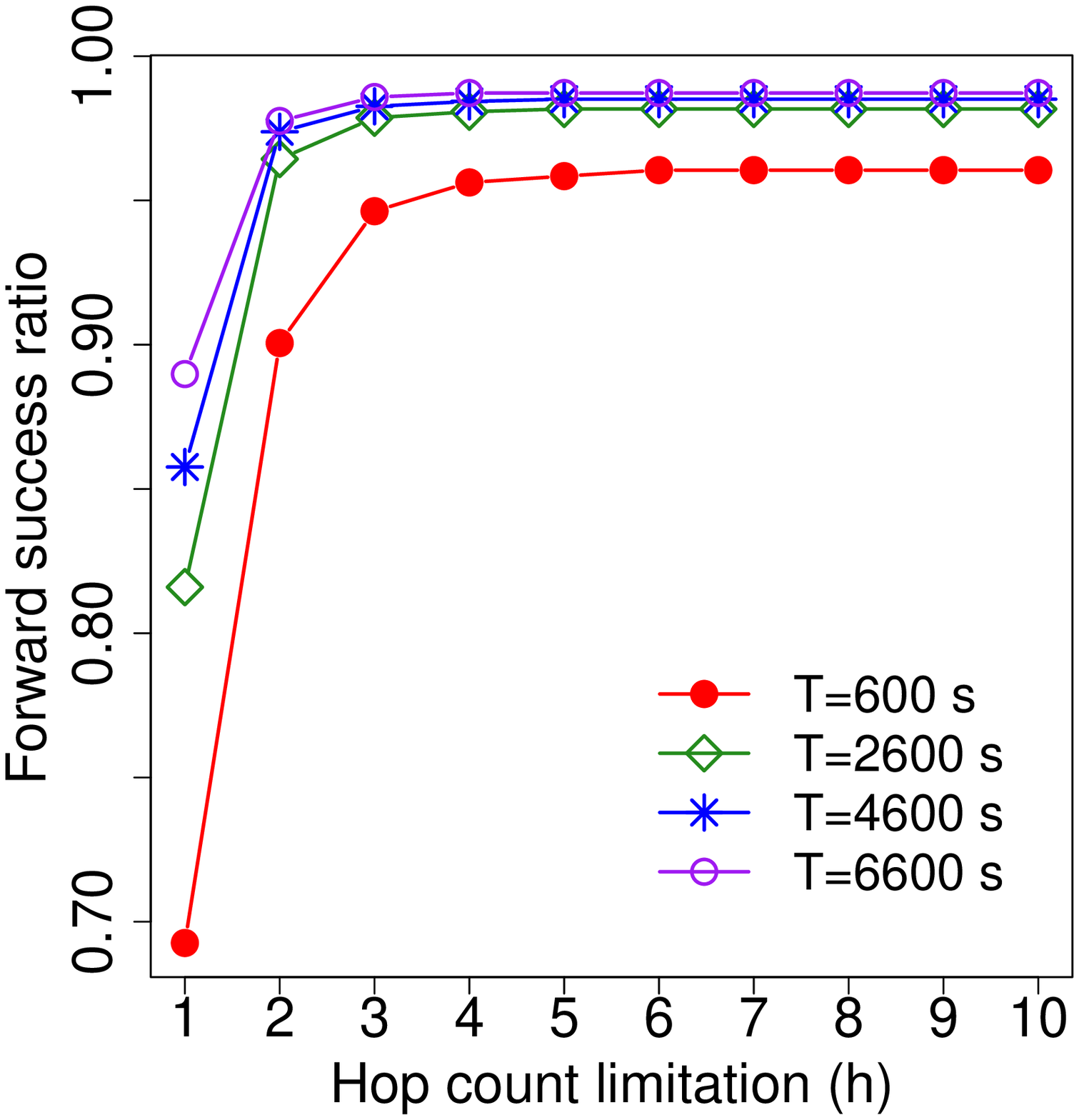}
                \label{fig:Infocom06_Ptagged_98_t05}}
     \subfigure[Infocom06, $T=600$ s.]{
                  \includegraphics[scale=0.22]{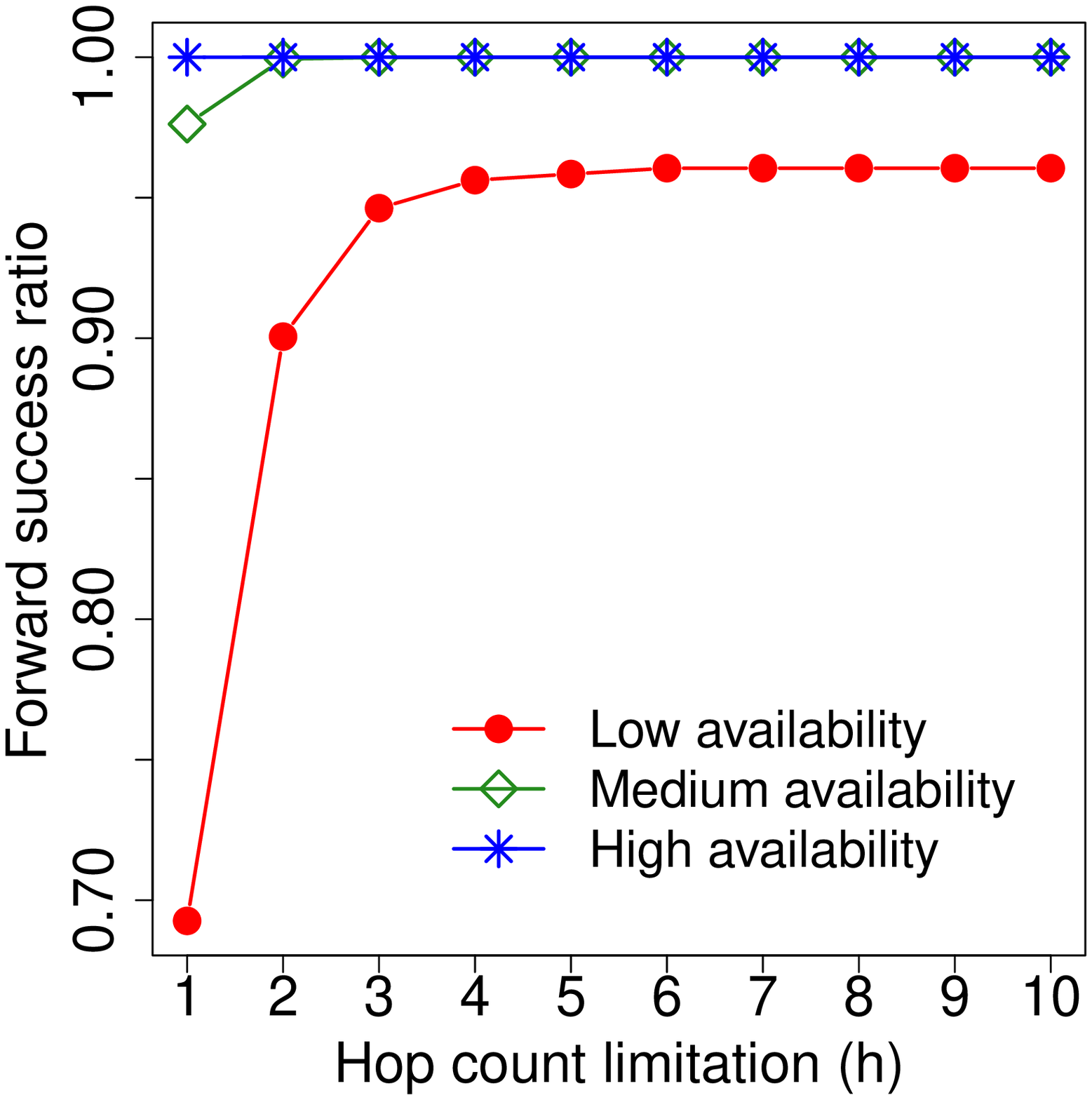}
                  \label{fig:Ptagged_98_t600}}
     \subfigure[HCS, $N_h(T)$.]{\includegraphics[scale=0.22]{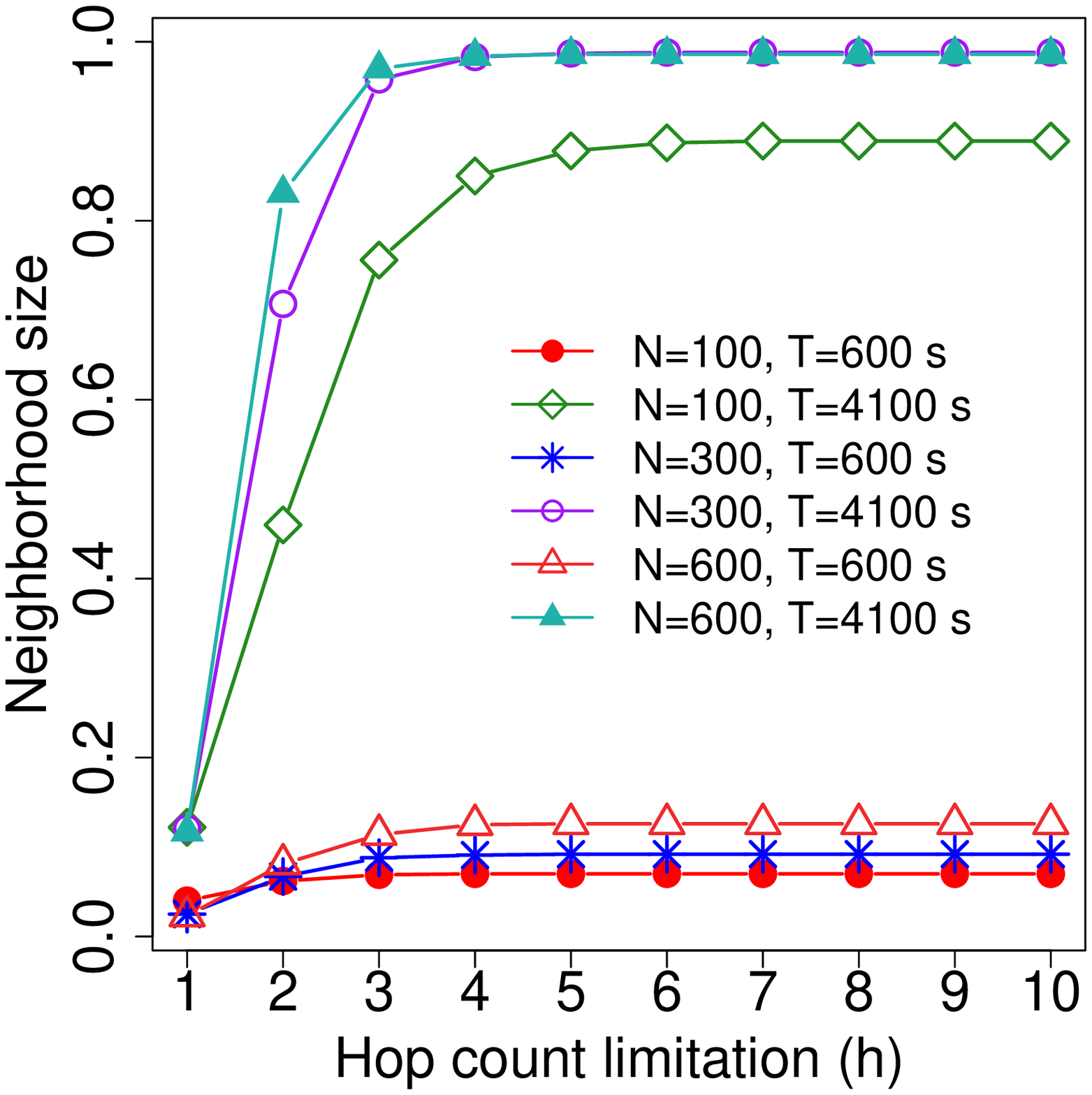}
                 \label{fig:density_vs_neighborhood}}
    \caption{(a) Growth of $h$-hop neighborhood for Infocom06 under various tolerated waiting time ($T$)   and (b) corresponding forward success ratio for $\avail=0.05$. (c) Effect of content availability on $P_h(T)$ for Infocom06. (d) Growth of $h$-hop neighborhood for HCS under various $N$ and $T$. 
            \label{fig:Neighborhood}}
     \end{figure}

Fig.~\ref{fig:Neighborhood} illustrates the change in $\Ndis_h(T)$ represented as fraction of the network size for Infocom06 for various tolerated waiting time $T$ and content availability $\avail$. For each $T$, we plot $\Ndis_h(T)$ and corresponding $P_h(T)$. 
From Fig.~\ref{fig:Infocom06NodeUnionByHop_98}, %
 we can see the significant growth of $\Ndis_h(T)$ at the second hop for all settings. 
 While further hops introduce some improvements, we observe the existence of a \textit{saturation point}~\cite{hoque2014efficient}. 
 After this point, the change in $\Ndis_h$ is marginal either because all nodes are already covered in the neighborhood or higher $h$ cannot help anymore without increasing $T$. 
 
We present the resulting $P_h$ for low content availability in Fig.\ref{fig:Infocom06_Ptagged_98_t05}.
 As expected, the second hop provides the highest performance gain compared to the previous hop ($\Delta P_{h,h+1}$) as a reflection of highest $\Delta \Ndis_{h,h+1}$. As Infocom06 has good connectivity, almost all queries reach one of the tagged nodes after $h\approx 4$.

We present the effect of content availability for short $T$ in Fig.~\ref{fig:Ptagged_98_t600}.
Regarding the effect of content availability, \textit{we observe that search for a rare content item, i.e., low $\avail$, benefits more from increasing $h$ compared to highly available content items}. When many nodes are tagged, $n_s$ meets one of these after some time. However, if the probability of meeting a tagged node is fairly low, using additional hops exploiting the mobility of the encountered nodes and spreading the query further is a better way of searching. A smart search algorithm can keep track of the content availability via message exchanges during encounters and can adjust the hop count depending on the observed availability of the content. For low and medium content availability, \textit{the highest benefit is obtained at the second hop}. For a content item which is stored by a significant fraction of nodes, even a single hop search may retrieve the sought content.

Fig.~\ref{fig:density_vs_neighborhood} illustrates the growth of $N_h(T)$ under various network size $N$ and $T$. 
We use our synthetic model HCS by setting $N=\{100,300,600\}$ nodes to observe the effect of network density on $N_h(T)$.
For this particular setting, the clustering of lines based on $T$ shows that time restriction is more dominant factor in determining $N_h(T)$ compared to $N$.
As expected, $N_h(T)$ is higher for higher $N$. However, all settings exhibit the same growth trend with 
increasing $h$.

\begin{figure}
     \centering
     \subfigure[Infocom06, fair-benefit.]{\includegraphics[scale=0.223]{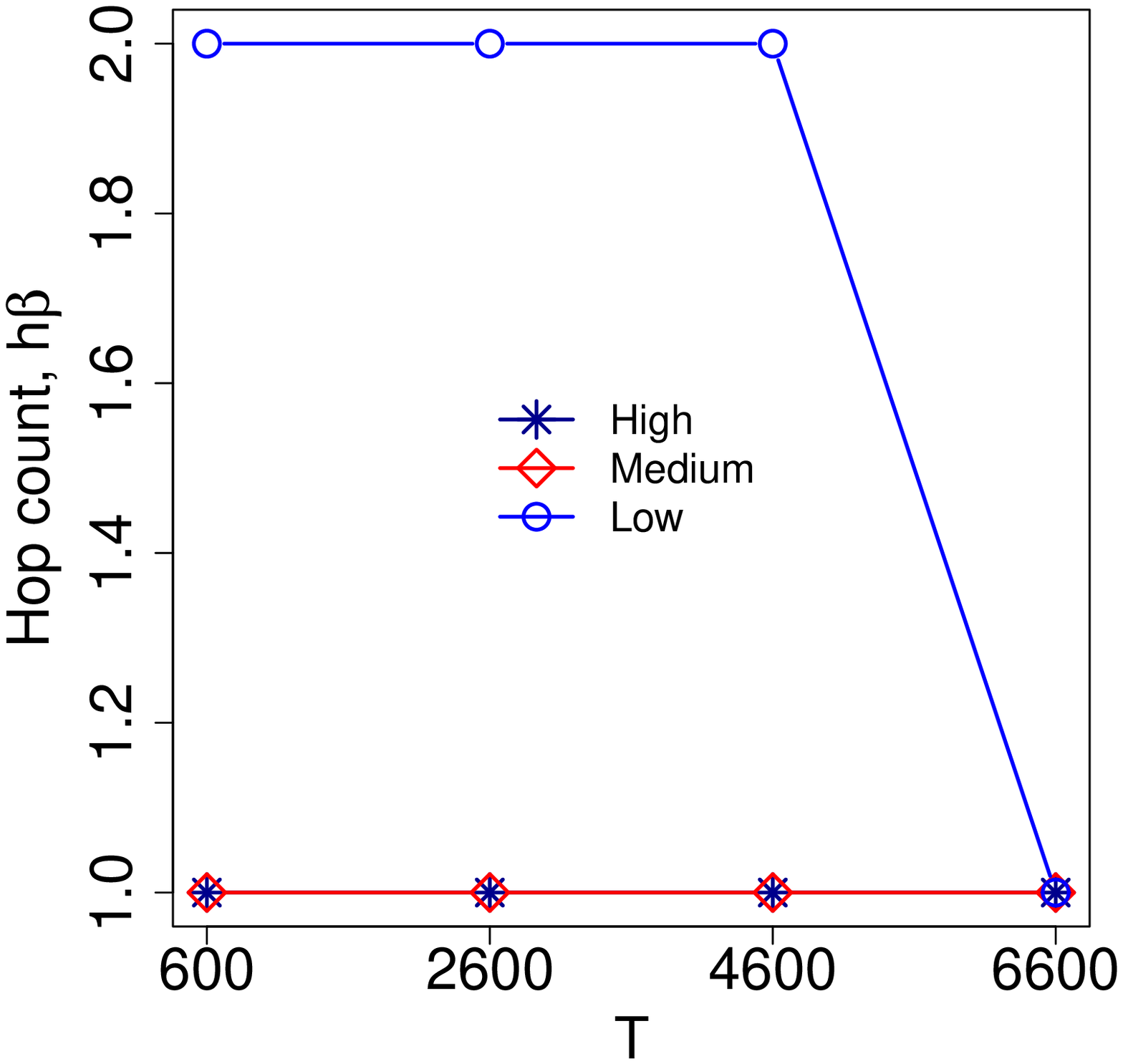}
     \label{fig:InfocomNoBenefitPoints_98_benefit5e-04Fair}}
     \subfigure[Infocom06, any-benefit.]{\includegraphics[scale=0.223]{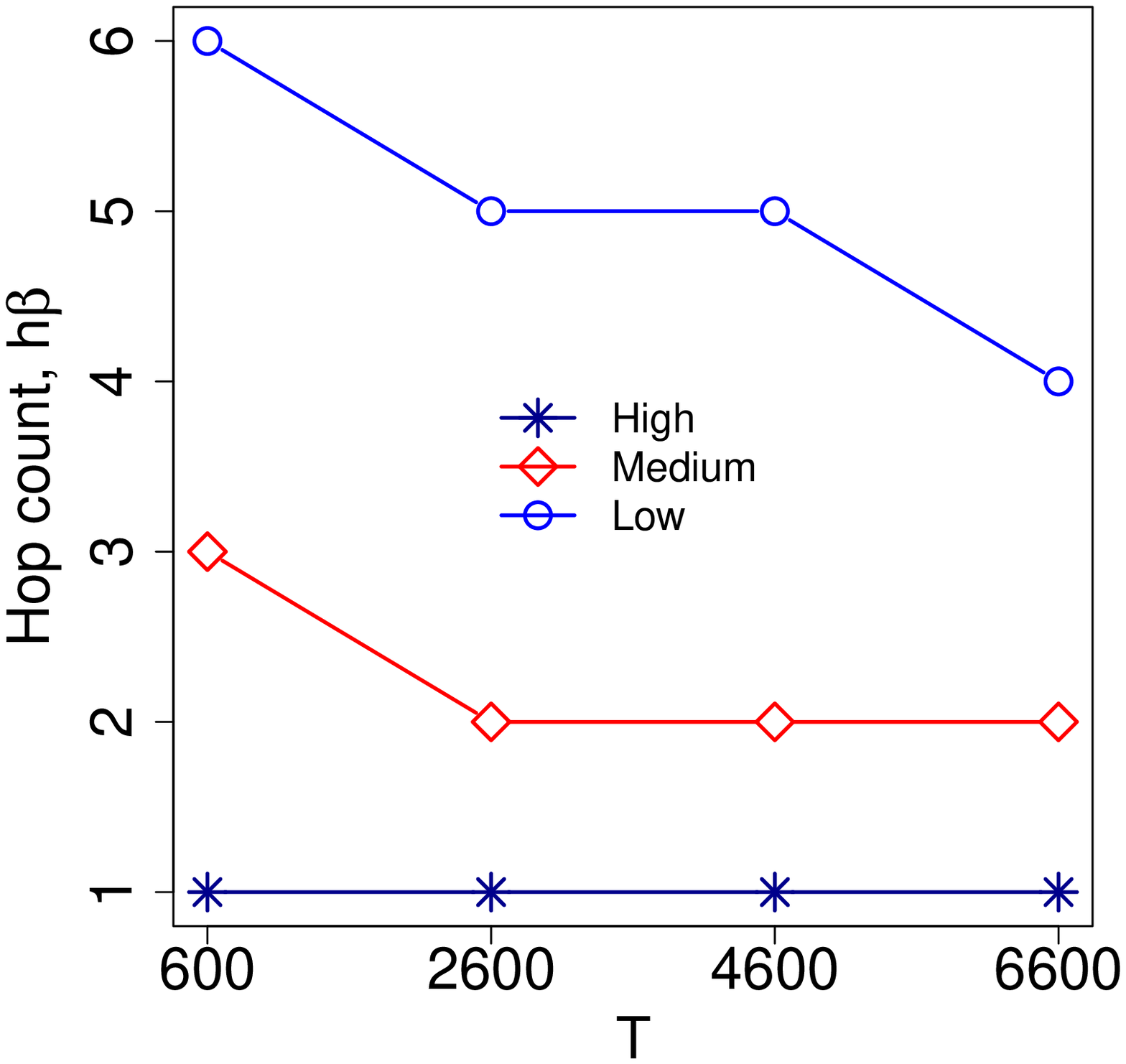}
     \label{fig:InfocomNoBenefitPoints_98_benefit5e-04Any}} 
     \subfigure[HCS, fair-benefit.]{\includegraphics[scale=0.223]{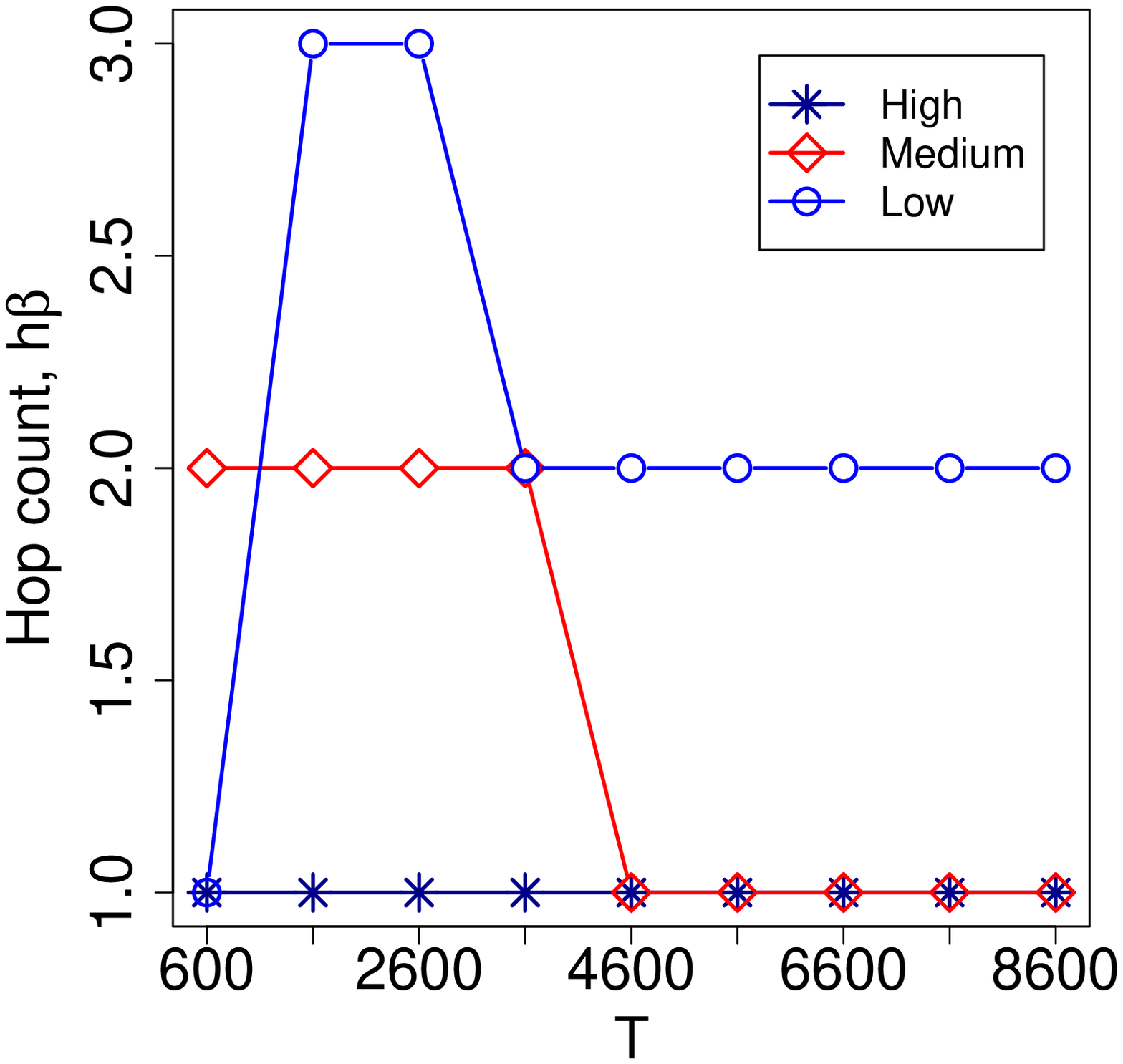}\label{fig:HelsinkiFair}}
     \subfigure[HCS, any-benefit.]{\includegraphics[scale=0.223]{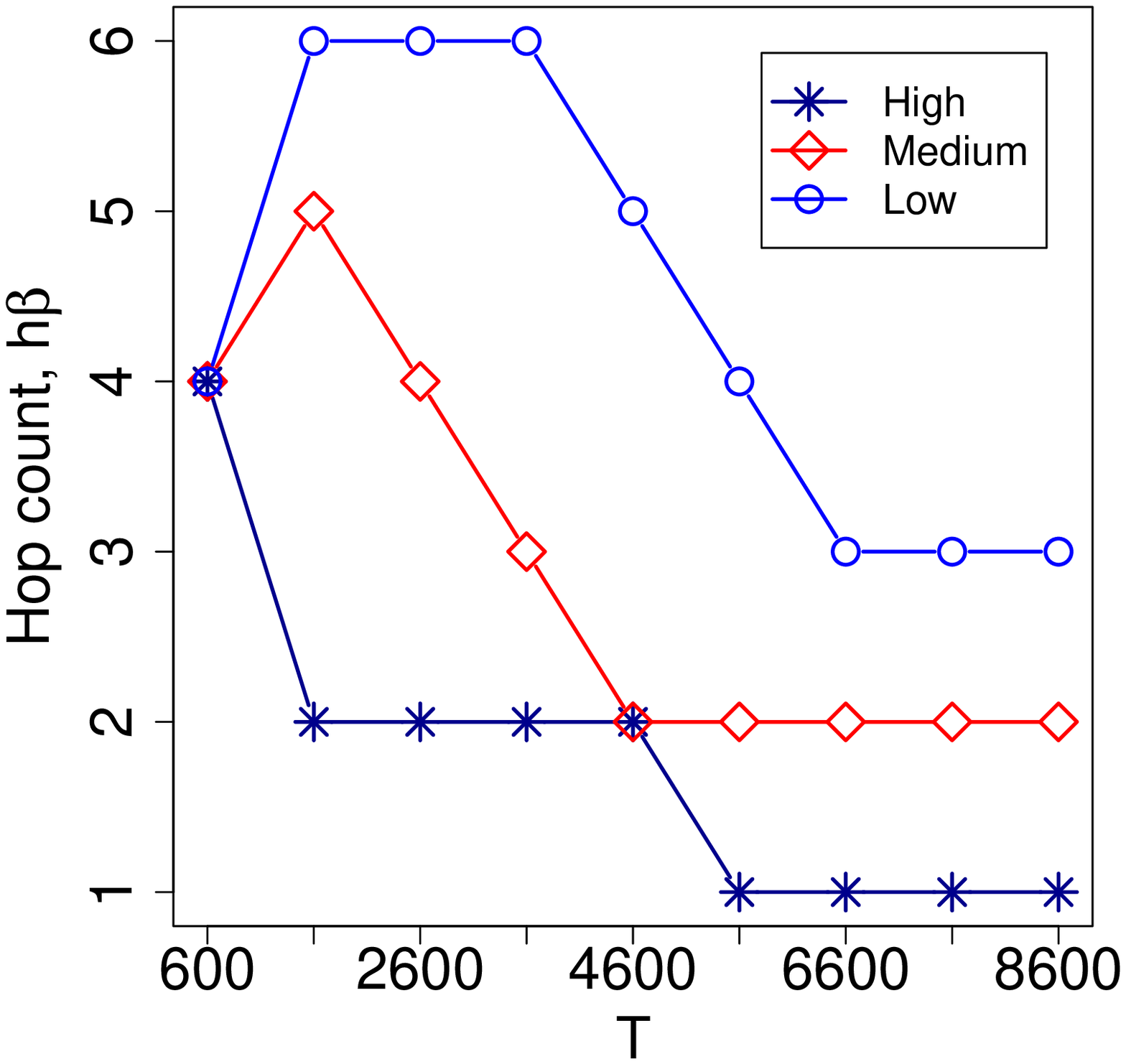}
     \label{fig:HelsinkiAny}}
     \caption{Fair- and any-benefit hops.\label{fig:AllBeta}}
    \end{figure}

\subsection{Fair-benefit and Any-benefit Hops}
Fig.~\ref{fig:AllBeta} illustrates the change in $h_{\beta}$ with increasing $T$ for $\beta\in\{0.0005, 0.1\}$ which represent any-benefit and fair-benefit hops, respectively. 
As stated before, Infocom06 has good connectivity among nodes as conference takes place in a closed area and nodes share the same schedule (e.g., coffee breaks and sessions). As a result of this good connectivity,  first one or two hops provide almost all the benefits of multi-hop search (Fig.~\ref{fig:InfocomNoBenefitPoints_98_benefit5e-04Fair}). 
However, Fig.\ref{fig:InfocomNoBenefitPoints_98_benefit5e-04Any} shows that the optimal hop in terms of the highest $P_h$ is achieved at higher $h\leqslant 6$. 

For HCS scenario that represents a network of urban scale, the effect of increasing $T$ is a bit different. Fig.~\ref{fig:HelsinkiFair} shows that short and longer $T$ may have the same operating point in terms of $h_{\beta}$. For low $T$, the benefit of increasing $h$ diminishes due to the limited number of encounters and the resulting small set of discovered nodes, whereas for long $T$ almost all nodes are discovered without requiring any further hops. The lower $h_{\beta}$ with increasing $T$ can also be explained by ``\textit{shrinking diameter}'' phenomenon, which states that the average distance between two nodes decreases over time as networks grow~\cite{leskovec2005graphs}. Indeed, the diameter of %
the message search tree
gets shorter over time and leads to a smaller hop distance between the searching node and a tagged node. Note the decreasing \textit{any-benefit} hop in Fig.~\ref{fig:HelsinkiAny}. For example, searching for a content item with medium availability achieves the highest performance at $h_\beta=5$ for $T=1600$ s, while it decreases gradually to $h_\beta=2$ with increasing $T$.

\subsection{Time to Forward Path Completion}

Table~\ref{tab:SearchTime_MarkovChain} shows the search time $T_h$ which is 
obtained by solving CTMC introduced in Section~\ref{sec:TTsearchcompletion} and 
 $\tilde{T}_h$ given by (\ref{eq:TmhApproxm1}). 
 We normalize every value by the maximum search time of each setting.
In the table, we also list the approximation errors. 
First thing to note is the drastic decrease in search time at the second hop. 
In agreement with our conclusions from $P_h$, we see that second hop speeds the search significantly resulting in approximately $80\%$ shorter search time for the low availability, $60\%$ for the medium, and $50\%$ decrease for the high availability scenario. 
As $P_h$, the most significant improvement in search time occurs for low content availability. Our approximation also exhibits exactly the same behaviour in terms of the change in the search time. As the error row shows, it deviates from the expected search time to some degree: $32\%$ under-estimation to $8\%$ overestimation. 

\begin{table} [!htb]
 \centering 
% \small
 \caption{Search time and its approximation.}
 \renewcommand{\arraystretch}{1}
 \begin{tabular}{|llccccc|}\hline 
$\avail$& Time&$h=1$ & $h=2$ & $h=3$& $h=4$ & $h=5$\\ \hline %
\multirow{3}{*}{Low}&	$T_h$ & 1 &	0.21 &	0.13&	0.12& 	0.11 \\  
  &	$\tilde{T}_h$ &1 &	0.14&	0.12 &	0.11 &	0.11  \\  
 &	Error &0 & -0.32&	-0.08&	-0.03&	-0.03   \\ \hline
\multirow{3}{*}{Medium} &	$T_h$& 1 & 0.40 & 0.33 &0.31&	0.31 \\  
 &	$\tilde{T}_h$ & 1	& 0.39& 	0.35& 	0.33& 	0.33 \\  
 &Error &	0 &	-0.01&	0.07&	0.06&	0.05\\ \hline
 \multirow{3}{*}{High} &	$T_h$ &1 & 0.51 &	0.46&	0.45&	0.45  \\  
 &	$\tilde{T}_h$ & 1& 	0.54& 	0.49& 	0.47&	0.46\\ 
 &Error & 0	& 0.05 &	0.08&	0.06&	0.04\\ \hline
\end{tabular}
 \label{tab:SearchTime_MarkovChain}
 \end{table}

\section{Complete Search}\label{sec:Sims}
In Sec.\ref{sec:NumericalAnalysis}, we show that the first few hops yield the highest benefit in terms of forward path success ratio. In this section, we explore if this trend holds for the whole search, i.e., forward and return path. To this end, we 
carry out a wide range of simulations 
to gain insight to the following aspects:
(i) the fair-benefit hops for various $T$ and $\avail$ settings,   
(ii) average temporal and hop distance to/from content,   
(iii) characteristics of the return path, and  
(iv) search cost.

Note that return path also applies hop-limited routing as well as the forward path.
We first assume that an oracle %
 stops the dissemination of a completed search immediately, and then relax this assumption in a scheme where nodes are informed about search state via control messages.  
 Using the ONE simulator~\cite{keranen-theone}, we design an opportunistic network consisting of $N=98$ nodes for Infocom06 and $N=496$ for Cabspotting scenario. 5000 content items are distributed across nodes according to the availability ratios: for example, a content item with $\alpha=0.4$ is randomly assigned to $40\%$ of the nodes. This assignment is static during the course of the simulation: nodes do not generate new content items and nodes having received a copy during operation will not respond to
        requests.
Every query interval, a random node initiates a query for a uniformly chosen
 content item. The query generation interval is uniformly distributed
in [10,20]\,s. %
 We simulate the complete system 
during which approximately 23000 queries are generated. 

We use 20\,m radio range for Infocom06 and 40\,m for Cabspotting, infinite transmission capacity (i.e., all packets can be exchanged at an encounter), and 100\,MB of storage capacity; message are 15\,KB in size. To obtain an upper performance bound, we implement \textit{epidemic search} (EPID) which does not impose any hop limitations. 
In the following, we refer to $h$-hop search as HOP.

\subsection{Fair-benefit and Any-benefit Hops}\label{sec:Verif}

 \begin{figure}
 \centering
 \subfigure[$P_s$, Infocom06.]{\includegraphics[scale=0.23]{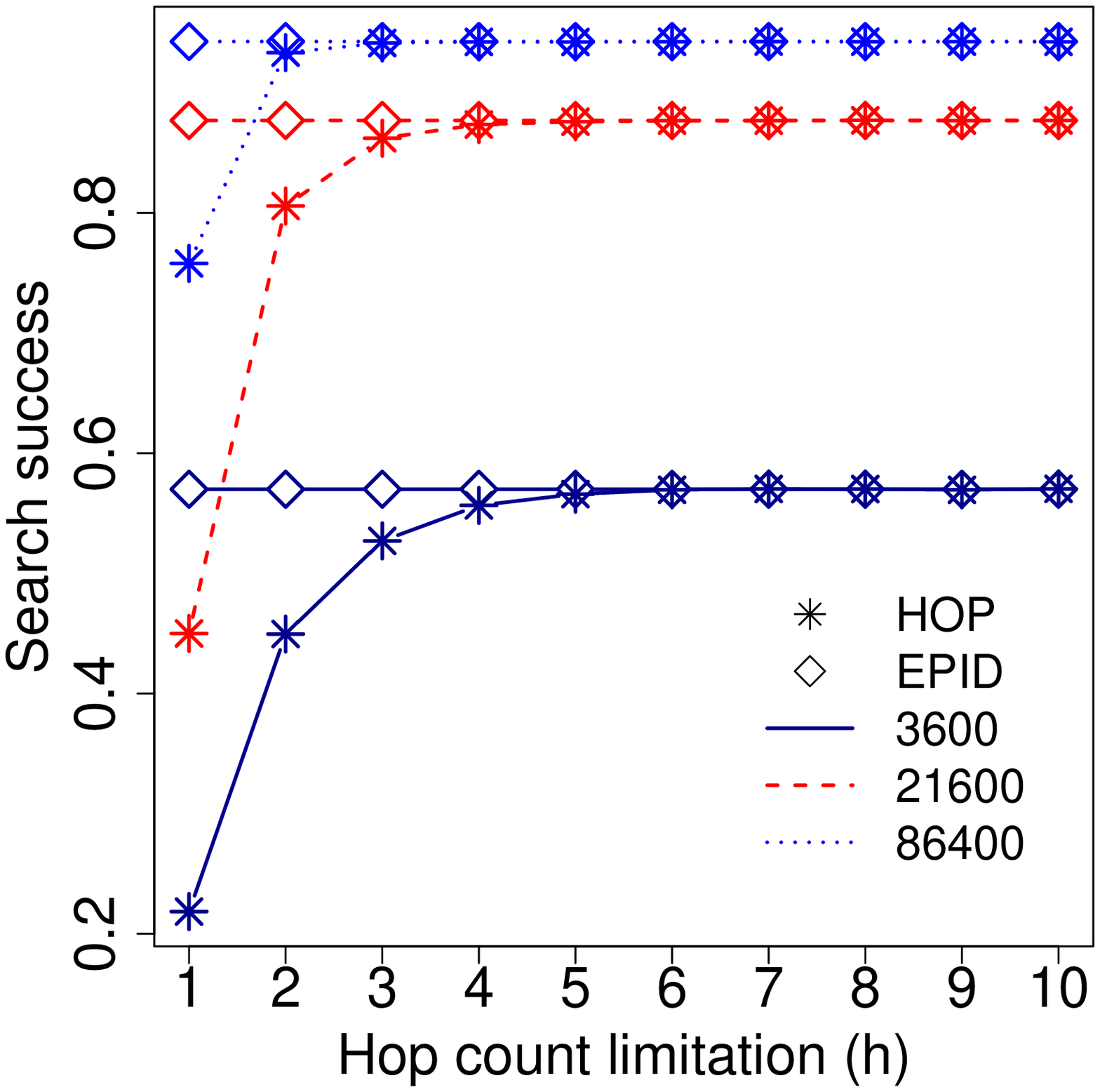}\label{fig:all_ttl_HOP_N98_05_Searchsuccess}}
 \subfigure[$P_{h}$, Infocom06.]{\includegraphics[scale=0.23]{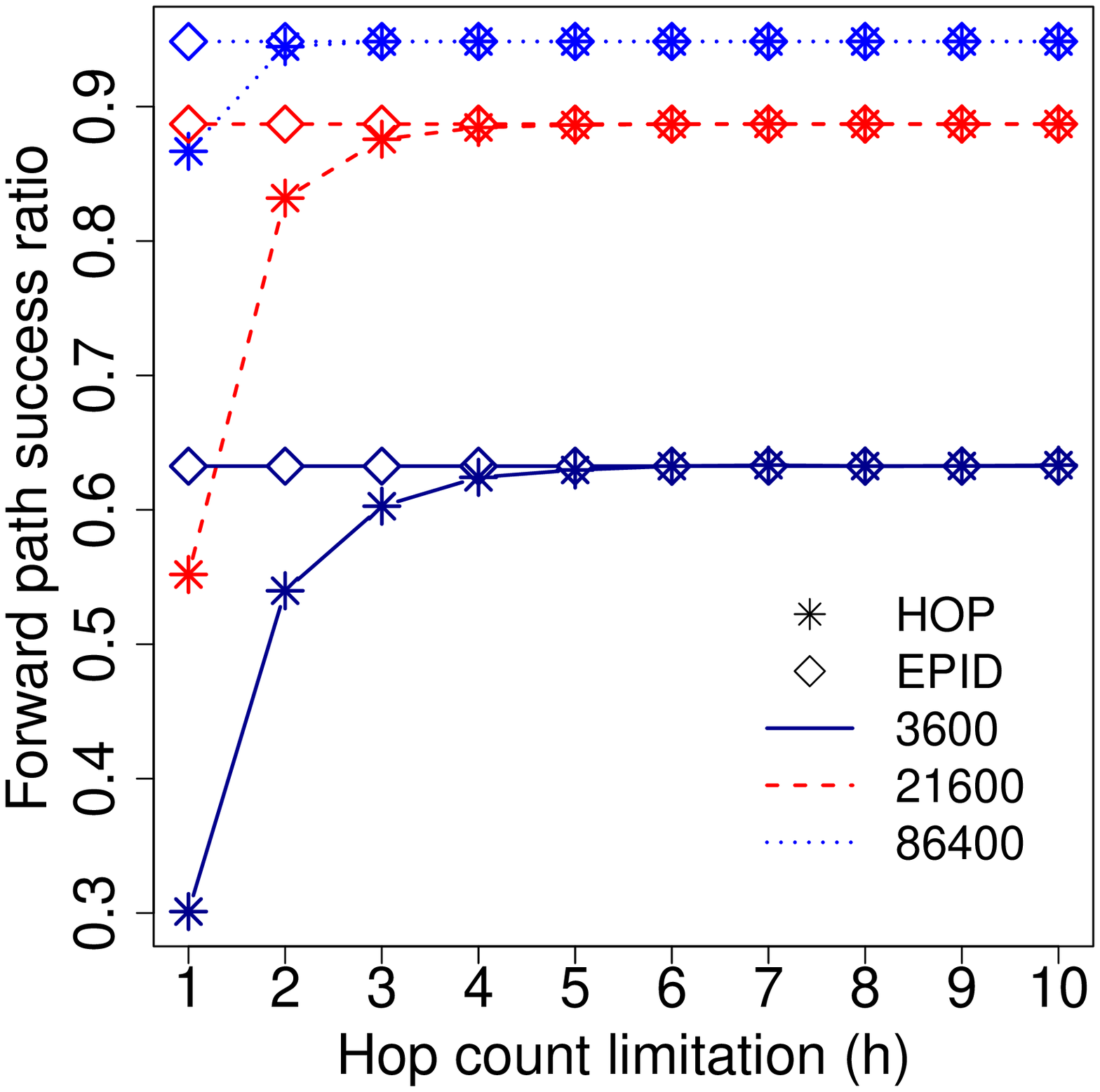}\label{fig:all_ttl_HOP_N98_05_Forwardpathsuccessratio}}
 \subfigure[$P_s$, Cabspotting.]{\includegraphics[scale=0.23]{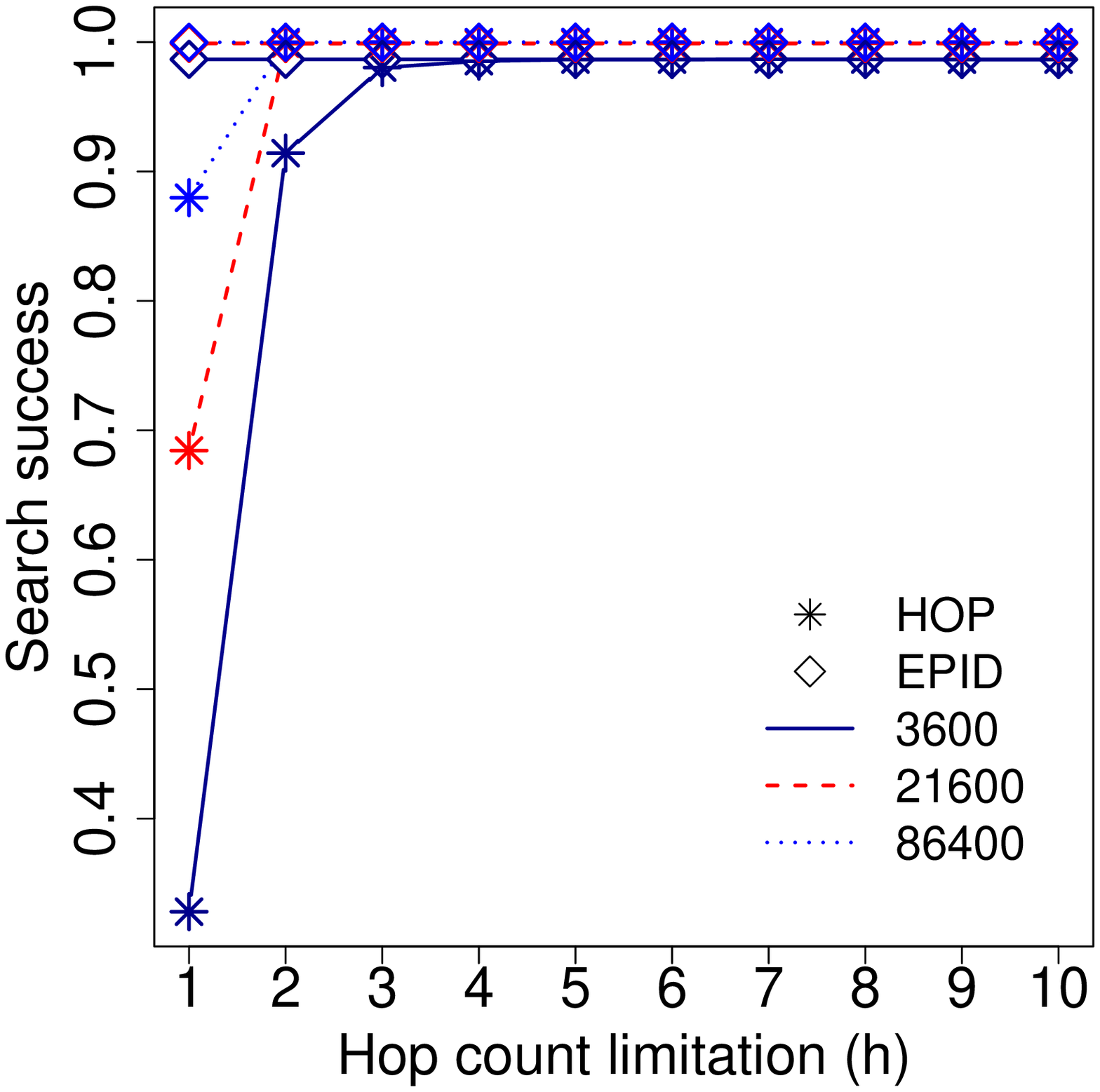}\label{fig:all_ttl_HOP_N496_05_Searchsuccess}}
 \subfigure[$P_{h}$, Cabspotting.]{\includegraphics[scale=0.23]{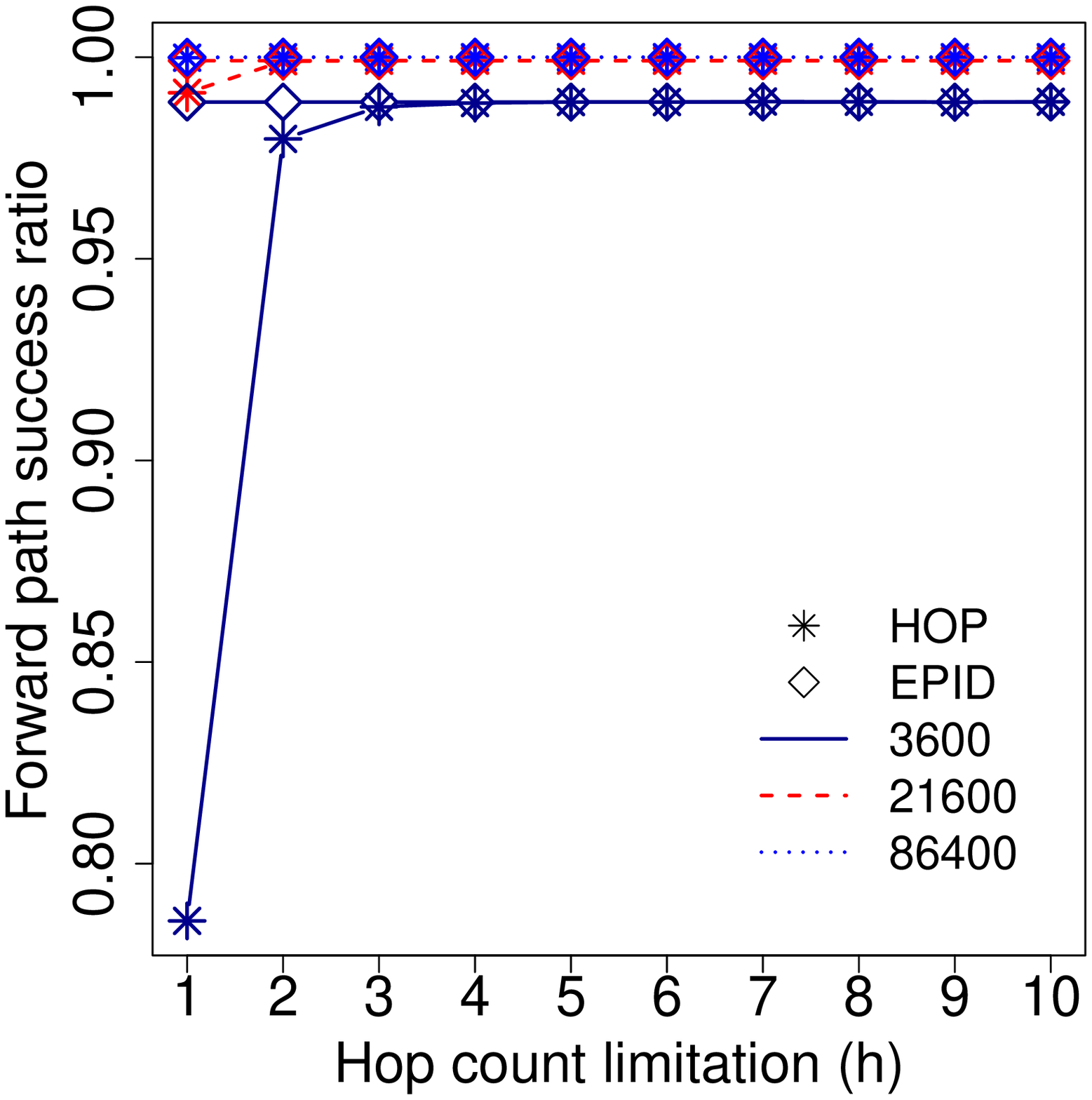}\label{fig:all_ttl_HOP_N496_05_Forwardpathsuccessratio}}
 \caption{Effect of tolerated waiting time, $\avail=0.05$.\label{fig:Sim_allTTL_HOP_Infocom06_SF}}
 \end{figure}

 \begin{figure*}
 \centering
 \subfigure[$P_s$, Infocom06.]{\includegraphics[scale=0.23]{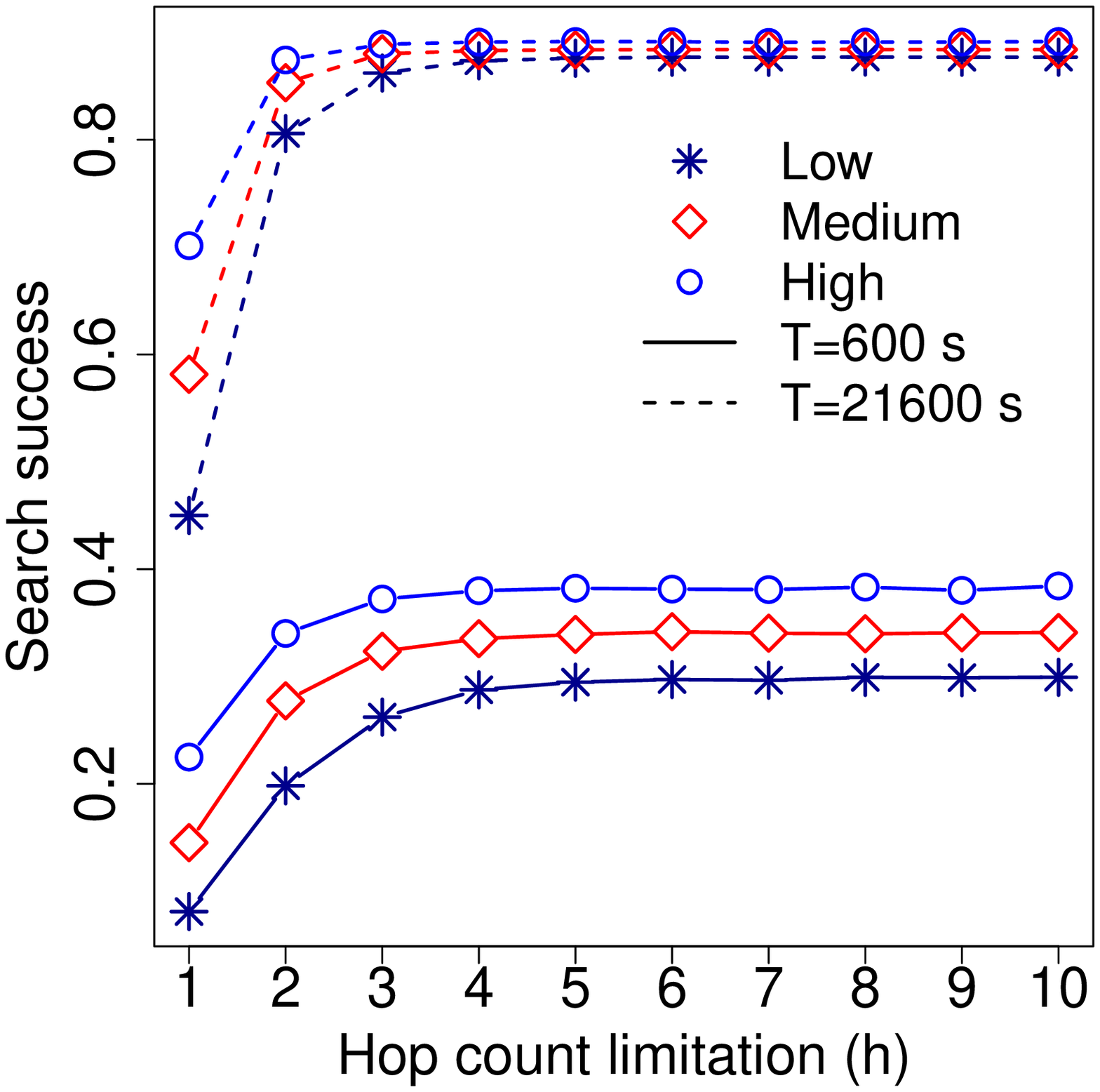}\label{fig:all_ca_HOP_N98_05_Searchsuccess}}
 \subfigure[$P_s$, Cabspotting.]{\includegraphics[scale=0.23]{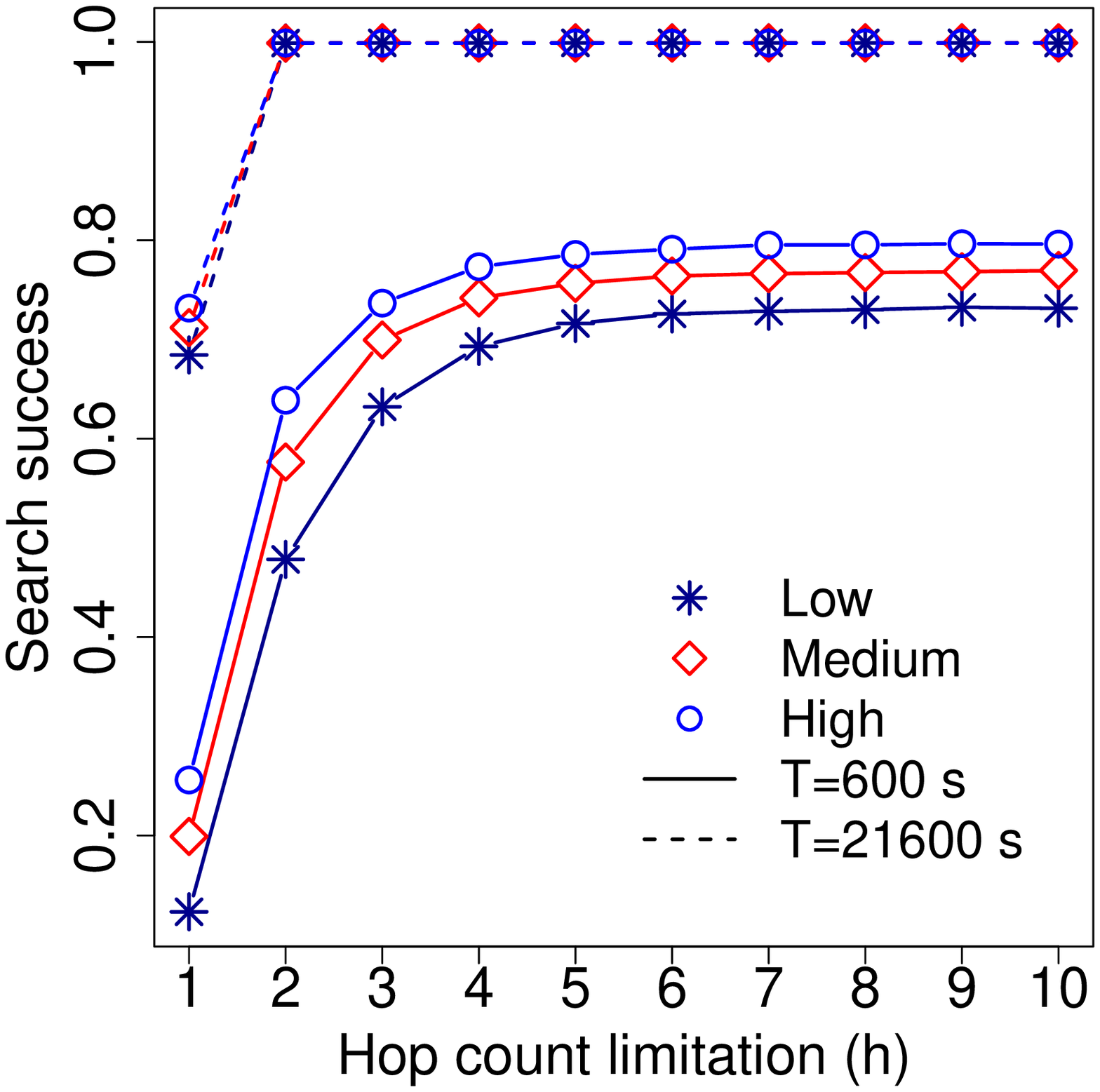}\label{fig:all_ca_HOP_N496_600_Searchsuccess}}
 \subfigure[Hop count, Infocom06.]{\includegraphics[scale=0.23]{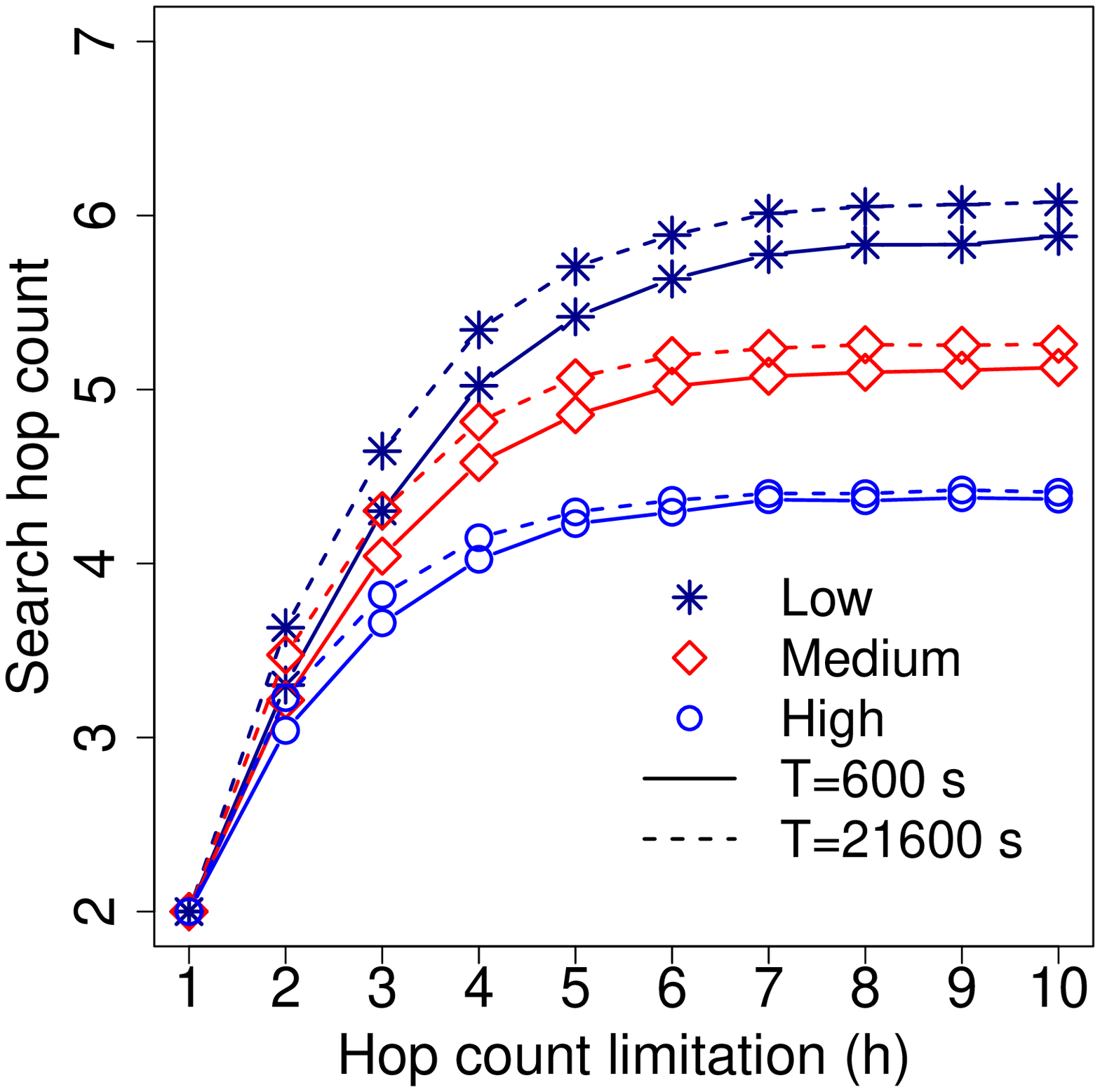}\label{fig:all_ca_HOP_N98_600_Searchhopcount}}
 \subfigure[Hop count, Cabspotting.]{\includegraphics[scale=0.23]{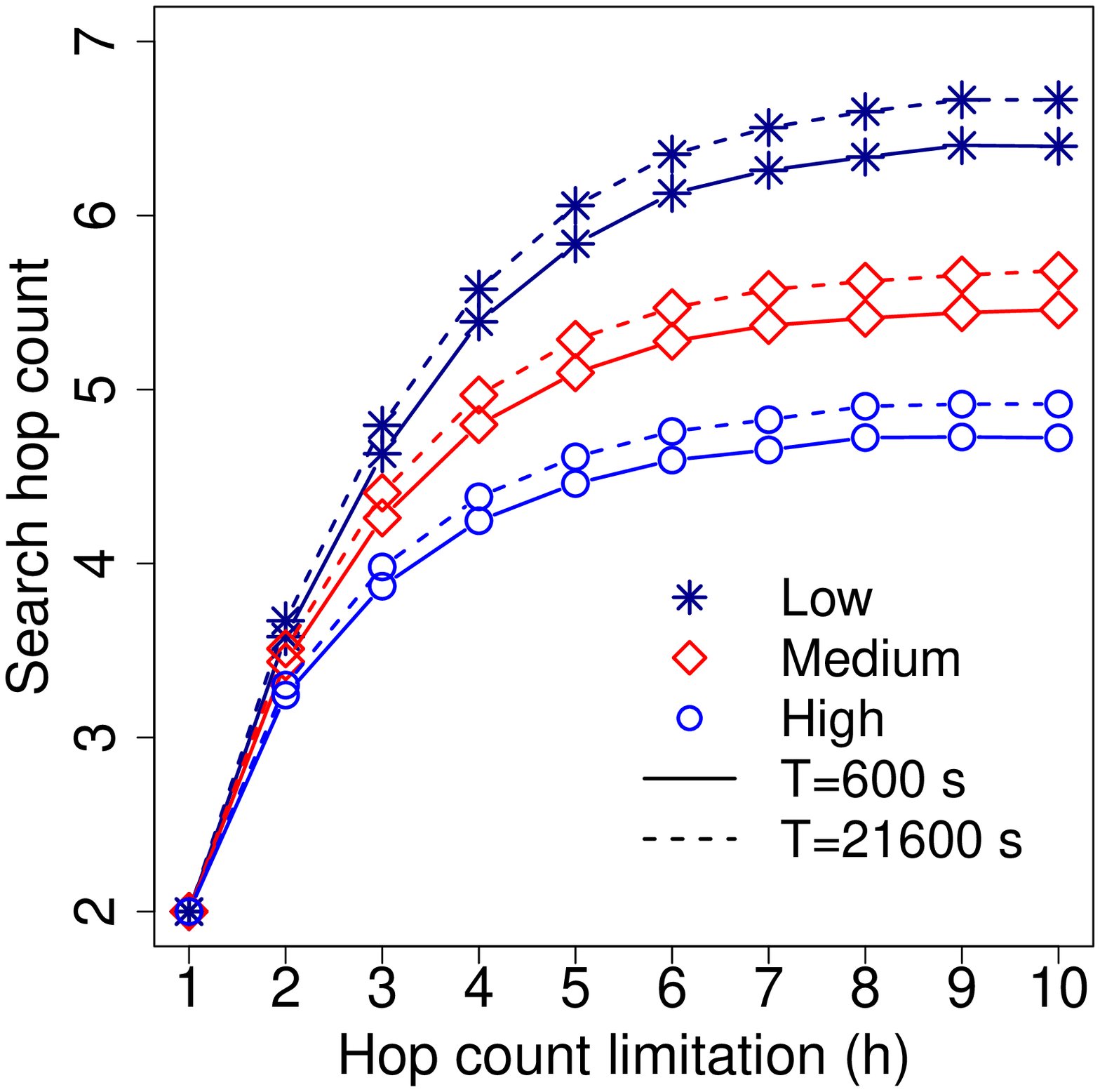}\label{fig:all_ca_HOP_N496_600_Searchhopcount}}
 \caption{Effect of content availability under  $T=600$ and $T=21600$ s.\label{fig:Sim_allca_HOP_Infocom06}}
 \end{figure*}

We analyse the effect of $h$ on the search success ratio~($P_{s}$, the ratio
of queries retrieving a response) and forward path success ratio~($P_{h}$). Fig.~\ref{fig:Sim_allTTL_HOP_Infocom06_SF} depicts $P_{s}$ and $P_{h}$ for HOP and EPID under various settings: $T=\{3600, 21600, 86400\}$ s, all availabilities, and for Infocom06 and Cabspotting scenarios. Total tolerated search time~(i.e., forward and the return path) is set to $2T$.

As Fig.~\ref{fig:Sim_allTTL_HOP_Infocom06_SF} shows, the highest $\Delta P_{h,h+1}$ is at the second hop for all $T$. However, it takes more hops until HOP achieves the same success ratio as EPID: $h=5$ for $T=3600$ s, $h=4$ for $T=21600$ s, and $h=3$ for $T=86400$ s. Cabspotting scenario achieves higher success ratio under the same time limitations compared with Infocom06. Although Infocom06 represents a small area and closed group scenario, we observe more contacts in Cabspotting setting. This may be attributed to higher transmission range as well as higher mobility of the nodes. 

If we consider only the forward path as we did previously
in Fig.~\ref{fig:Infocom06_Ptagged_98_t05}, we observe in Fig~\ref{fig:all_ttl_HOP_N98_05_Forwardpathsuccessratio} that the second hop provides almost all the benefits for longer waiting times. For more strict $T$, we see that the search success ratio is below 1 meaning that all benefits of multi-hop routing is exploited. Under this setting, the search performance cannot be improved without increasing $T$. Having visited $P_s$ and $P_h$ figures for Infocom06, we conclude that although the target content is discovered at the second hop with high probability, it does not necessarily ensure that the return path completes in two hops. As such, return path is similar to a search for the content with $\alpha = 1/N$, i.e., a scarce content item. Hence, it may require further hops. For Cabspotting, two to three hops yield the same performance as EPID.

Figs.~\ref{fig:all_ca_HOP_N98_05_Searchsuccess} and ~\ref{fig:all_ca_HOP_N496_600_Searchsuccess} demonstrate the effect of content availability on the whole search success for $T=600$ s and $T=21600$ s. As we see in the figures, $P_s$ values are clustered based on $T$ for both Infocom06 and Cabspotting. This behaviour can be interpreted as the tolerated waiting time being more dominant factor in determining the search performance rather than the content availability under the considered settings. Regarding $h_{\beta}$, fair-benefit hops are around 2-3 hops whereas any-benefit hops vary from 2 to 6 hops. Finally, Figs.~\ref{fig:all_ca_HOP_N98_600_Searchhopcount} and ~\ref{fig:all_ca_HOP_N496_600_Searchhopcount} demonstrate the total hop count for the whole search route. In contrary to $P_s$, we observe a clustering according to content availability in the total search hop counts. While the tolerated waiting time has some effect on the average search path length, the effect of content availability is the dominating factor. Low availability content items are retrieved from more far-away nodes whereas items with higher availability are retrieved from closer nodes. Nevertheless, the total path lengths are much shorter than the imposed limit, which means that content is typically retrieved from ``nearby'' nodes.

\subsection{Effective Distance to/from Content}\label{sec:AvgDistance}
In the previous section, we presented experiment results for different tolerated waiting time settings and referred to them as \textit{short} or \textit{long} $T$. 
In fact, temporal dynamics of the network (e.g, average inter-contact time) determine whether a $T$ value is short or not. In this section, we aim to provide insights about the distance to content both in terms of number of hops and time. To this end, we focus on the forward paths under EPID and remove the time restriction (i.e., $T=\infty$). In general, the performance of an opportunistic network is governed by the routing protocol and the node mobility (among other factors). We eliminate the effect of the algorithm by using EPID. This allows us finding the average hop distance between a searching node and a tagged node. Similarly, as we avoid any routing effects, we obtain the shortest time it takes to reach a tagged node from a searching node. We refer to this time as \textit{temporal distance} to content. Let us define \textit{effective distance to content} as the maximum distance, be it hops or time, which ensures that 90$\%$ of the paths between a searching node and a tagged node is lower than this distance~\cite{leskovec2005graphs}. We define \textit{effective distance from content} similarly for the return path. 
 
 \begin{figure}[!htb]
 \centering
 \includegraphics[width=0.5\textwidth]
 {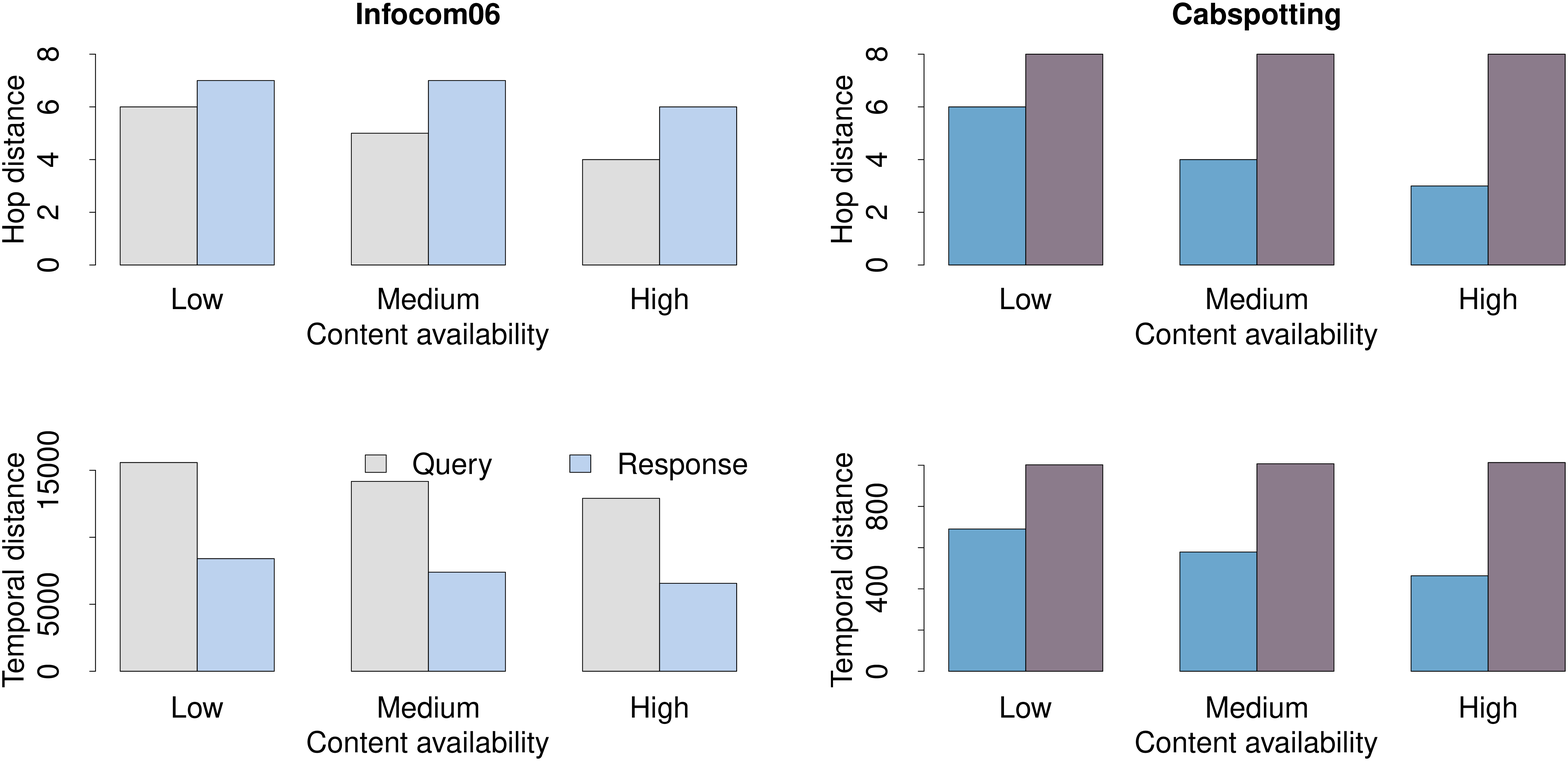}\label{fig:bar_hopdiameter_1}%
 \caption{Effective hop and temporal distance for query and return path for Infocom06 and Cabspotting. Left bars: query, right bars: response.\label{fig:Distances}}
 \end{figure}

In Fig.~\ref{fig:Distances}, we plot the effective hop and temporal distances for both the forward and return paths  under different content availabilities for Infocom06 and Cabspotting. While the effective distance to content~(i.e., forward path length) is inversely proportional to content availability, the effective distance from content (i.e., return path length) is only slightly affected by content availability. It is also noteworthy that the response hop distance is longer than the query distance for all settings. Regarding temporal distances in Fig.~\ref{fig:Distances}~(second row), Cabspotting has much shorter temporal distance compared to Infocom06. This difference explains higher success ratios achieved in Cabspotting compared with Infocom06 under the same $T$ in Fig.~\ref{fig:all_ttl_HOP_N496_05_Searchsuccess} vs. Fig.~\ref{fig:all_ttl_HOP_N98_05_Searchsuccess}. In Fig.~\ref{fig:Distances}~(bottom-left), effective temporal distance is shorter for the responses as opposed to higher hop counts observed in Fig.~\ref{fig:Distances}~(top-left). However, as we do not preserve the coupling between the query and response paths of the search, this result does not necessarily contradict with our claim that response path takes longer and is more challenging. In fact, responses that are still looking for the searching node are not accounted for, which may in turn lead to the shorter effective distance. Effective temporal distance is paramount for unveiling the conditions of feasibility of search. For example, setting $T=600$ s lead to very poor search performance for Infocom06 as shown in Fig.~\ref{fig:all_ca_HOP_N98_05_Searchsuccess} because the network evolves slower than this time. 

\subsection{Characteristics of the Return Path}\label{sec:Bwpath}
In the previous sections, we analyzed the query and response paths separately. One arising question is the relation between the forward and return paths: whether the latter depends on the former. In this section, we present the analysis of the query and response paths where query and response path coupling is preserved. We simulate EPID with a restriction on the search time. Using the completed queries, we calculate the Pearson correlation coefficient between the query and response hop~($\rho_{hop}$) as well as the query and response time~($\rho_{temp}$). We also find the ratio of the response hop to the query hop~($\gamma_{hop}$) as well as the ratio of response temporal path length to query temporal path length~($\gamma_{temp}$) to explore how challenging the return path is compared to the forward path.

\begin{table}[!htb]
 \centering 
 %\small
 \caption{Correlation between forward and return paths, Infocom06.}
 \renewcommand{\arraystretch}{1.1}
 \begin{tabular}{|cccccccc|}\hline 
 T&$\alpha$ & $\rho_{hop}$ & $\rho_{temp}$& $\gamma_{hop}$ & $\gamma_{temp}$ &$P_h$ & $P_s$ \\ \hline %
  \multirow{3}{*}{600}&	 Low &	 0.30 &	 0.36 &	 1.47 &	 2.23 &	 0.35 &	 0.30\\
  &	 Med. &	 0.29 &	 0.34 &	 1.72 &	 3.09 &	 0.42 &	 0.34\\
  &	 High &	 0.27 &	 0.32 &	 1.97 &	 4.02 &	 0.48 &	 0.38\\  \hline 
   \multirow{3}{*}{3600} &	 Low &	 0.35 &	 0.43 &	 1.4 &	 2.18 &	 0.63 &	 0.57\\
   &	 Med. &	 0.35 &	 0.38 &	 1.61 &	 2.98 &	 0.67 &	 0.61\\
  &	 High &	 0.33 &	 0.32 &	 1.85 &	 4.13 &	 0.70 &	 0.65\\  \hline 
  \multirow{3}{*}{86400}&	Low &	 0.33 &	 0.13 &	 1.39 &	 2.60 &	 0.95 &	 0.94\\
  &	 Med. &	 0.35 &	 0.13 &	 1.62 &	 3.52 &	 0.95 &	 0.94\\
  &	 High &	 0.35 &	 0.12 &	 1.86 &	 4.50 &	 0.95 &	 0.95\\  \hline 
\end{tabular}
 \label{tab:correlationAnalysis}
 \end{table}
 
Table~\ref{tab:correlationAnalysis} summarises this analysis for Infocom06, which also agrees with the analysis for Cabspotting scenario. First, we do not observe any strong correlation between return and forward path lengths for any of the settings (i.e., $\rho_{hop}$ and $\rho_{temp}$ are around 0.1--0.4). This result may be conflicting with the intuition that the information found in nearby/far-away will also be routed back quickly/slowly. However, search process is more complicated due to the intertwined effects of mobility and restrictions on the total search time. For example, consider a forward path with a very large number of hops. Due to the remaining short time before the tolerated waiting time expires, search can only go a few hops towards the searching node. This leads to a long forward path with a very short return path which challenges the above-mentioned intuition. With higher content availability, the required number of forward hops decreases whereas the return path length seems to be barely affected. Hence, the corresponding $\gamma_{hop}$ and $\gamma_{temp}$ increase. For all scenarios, return path is on the average longer than the forward path. Using this observation, a tagged node receiving a query can set the time-to-live field of its response message longer than the received query's forward path time.

\subsection{Cost of Search}
Obviously, increasing hop count increases the neighborhood which in turn increases the chance of finding the sought content. However, the larger neighborhood should also be interpreted as larger number of replication, i.e., higher search cost. In fact, the neighborhood size represents the upper bound of number of replications for a search message if no other search stopping algorithm is in effect. In this section, we evaluate the cost of search considering two simple mechanisms that aim to keep replication much below the upper bound set by $N_h(T)$.

We consider three cases: (i) ORACLE: there is a central entity from which a node retrieves the global state of a message in its buffer and can ignore it if outdated, e.g. a completed search or a query already reaching a tagged node, (ii) EXCH: 
upon encounter, nodes exchange their local knowledge about search activities in the network, (iii) LOCAL: 
nodes exchange their knowledge \textit{only} on the shared messages. 
 Note that an ORACLE can be an entity in the cellular network and nodes can access it via a control channel. In fact, this communication with the cellular network is more costly~(e.g., energy) compared to the opportunistic communication. Nevertheless, this scenario serves as an optimal benchmark to assess the
performance of the other scenarios. EXCH pro-actively spreads
information about existing queries to all nodes opportunistically,
which may be considered as leaking information to nodes not involved
in search. LOCAL circumvents this by only using its local knowledge
and sharing information with peers only about queries that the other
node has already seen.\footnote{This is easy to implement even without
cryptographic methods by using a two-stage protocol, in which nodes first
forward the queries they are carrying and then share information about
those queries they received from their peer.}
\begin{figure} %
\centering
\subfigure[Query spread ratio.]{\includegraphics[scale=0.23]{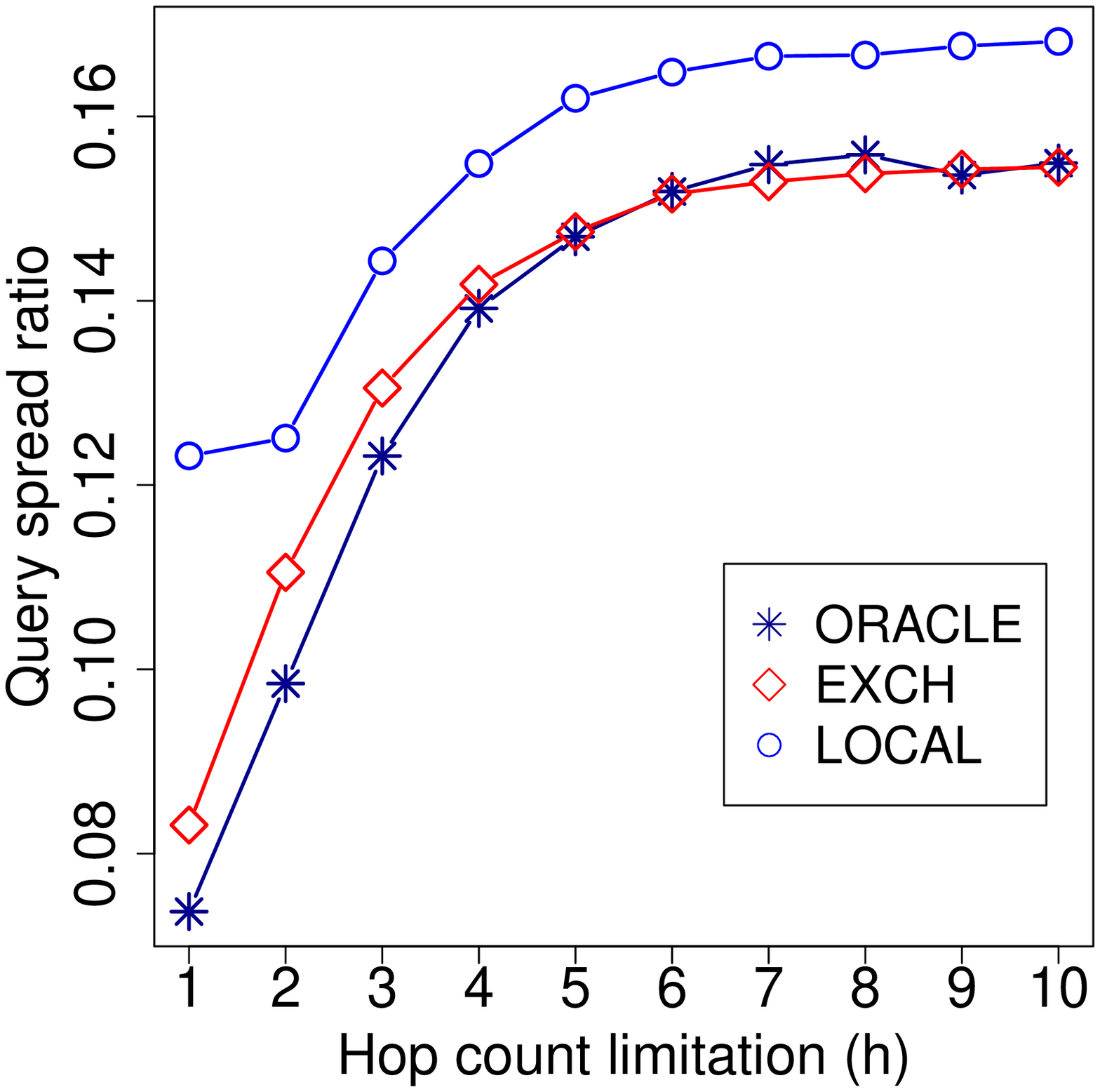}\label{fig:Queryspreadratio15_600_015}}
\subfigure[Search time.]{\includegraphics[scale=0.23]{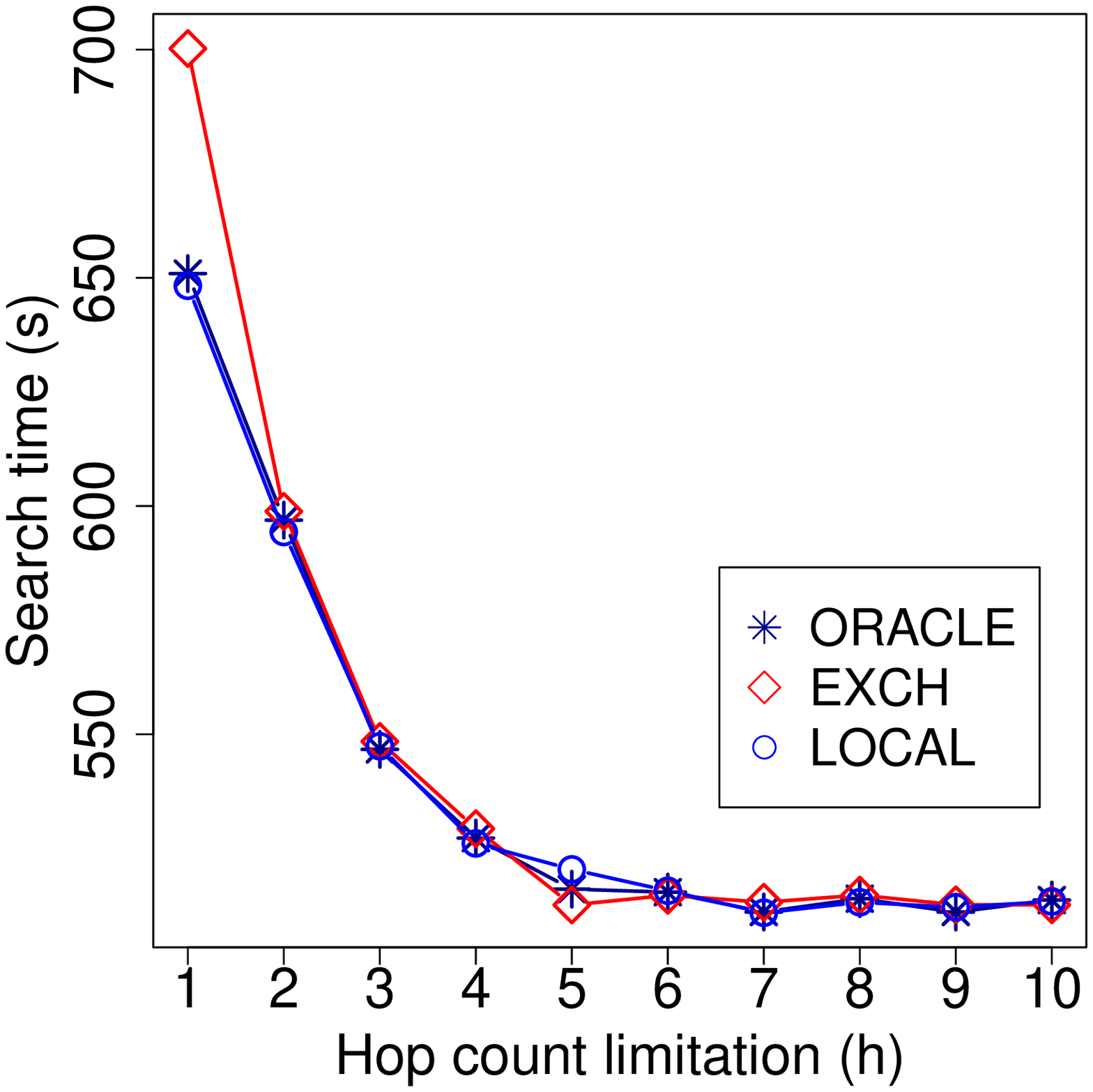}\label{fig:Search9_600_015}}
\caption{Hop count vs. search cost and time for Infocom06 ($\alpha$=0.15, $T=$600 s).\label{fig:costOraclevsNoOracle}}
\end{figure}

Fig.~\ref{fig:costOraclevsNoOracle} shows \textit{query spread ratio} which is defined as  i.e., the fraction of nodes that have seen this query, and search time. 
First, we should note that the resulting search success (not plotted) is almost the same under all schemes. Second, note the non-increasing query spread ratio for $h>5$. This result confirms that the network is a small-world network where all nodes get informed quickly about the search status and drop the outdated messages timely. Hence, even for larger $h$, nodes detect the outdated messages via local and shared knowledge. In Fig.~\ref{fig:Queryspreadratio15_600_015}, we observe that EXCH maintains the same performance as ORACLE,
whereas LOCAL results in more replication as fewer nodes are informed about the completed queries/responses. Nevertheless, because of the small network diameter, a higher $h$ does not result in an explosion of query spread in the network. 
Regarding search time, we observe substantial decrease in search time also for $h>2$ and $h<5$ in contrast to the vanishing benefits after second hop derived from our analytical model~(Table~\ref{tab:SearchTime_MarkovChain}). 
Search time tends to stabilise after $h\geqslant 5$. Hence, although several hops are sufficient in terms of search success, further hops (e.g., $h\approx 4$) can be considered for faster search.

\subsection{Discussion}
Our analysis shows that the two factors affecting the optimal hop count
are the content availability ($\alpha$) and the product of meeting
rate and tolerated waiting time ($\lambda T$). The latter is the
number of meetings before tolerated waiting time expires where $\lambda$ depends on network density and mobility model. If both
$\alpha$ and $\lambda T$ are low (scarce content and very few contacts
due to network sparsity or short search time), search
performance is expected to be low. However, it increases with increase
in either of these factors. If one of these factors is large, it is
sufficient to have a low hop count limit (e.g., two or three) to
obtain good performance. That is, when $\alpha\lambda T$ is large
relative to the expected number of tagged nodes met during the search
time, limiting the search to a few hops still achieves good 
results.  As $\alpha\lambda T$ decreases, the required hop limit to
maintain good search performance is larger. This allows devising adaptation when issuing search queries.  
Nodes can monitor the request and response rate for certain
(types of) content items and thus infer popularity and availability in
their area.\footnote{If nodes %
cache response contents opportunistically \cite{ott-aoc}, availability
  would grow with popularity.}  
  They can also assess the regional node
density and meeting rate $\lambda$ and thus determine the required 
hop count $h$ given $T$ or vice versa.
Moreover, nodes can monitor search performance and determine how well
their (region in a) network operates.
To improve performance, nodes can decide to increase the availability
of selected contents via active replication of the scarce content,
obviously trading off storage capacity and link capacity for
availability.  Such decision could be based entirely upon local
observations, but could also consider limited information exchange
with other nodes.

We believe that with the guidance of our analysis, a node can decide on the best hop count depending on the network density and the content availability that can both be derived from past observations by the node. Although we show that search cost stops increasing after a few hops, keeping $h$ low may be desirable if we interpret it as a measure of the social relation between two nodes (e.g., one hop as friends; two hop as friend-of-friend). 
In other words, lower $h$ can be interpreted as more {\em trustworthy} operation. Moreover, protocols involving lower number of relays are more scalable and energy-efficient~\cite{fan_delque_tvt2011}.

\section{Related Work}\label{sec:Related}

While the DTN literature has many proposals for message dissemination, which exploit the information about the network such as (estimated)
pairwise node contact
rates, %
node centrality, %
communities, %
and social ties, %
efficient mechanisms for content search remain largely unexplored.
In a sense, this is reasonable as search can be considered as a two-step message delivery: query routing on the forward path and the response routing on the return path. The forward path is less certain as the content providers are unknown. %
 In this regard, the return path is less challenging as the target node (and a recent path to reach it) is already known. However, routing the response looks for a particular node, whereas the forward search is for a \textit{subset of nodes} whose cardinality is proportional to the content availability.
 Thus, search requires special treatment rather than being an extension of the message dissemination.

Two questions for an efficient search are (i) which nodes may have the content~\cite{bayhan2013seeker} and (ii) when to stop a search~\cite{pitkanen2009searching}. %
The former question requires assessment of each node in terms of its potential of being a provider for the sought content. For example, \textit{seeker assisted search} (SAS)~\cite{bayhan2013seeker} estimates the nodes in the same community to have higher likelihood of holding the content as people in the same community might have already retrieved the content.
Given that people sharing a common interest come together at a certain \leftq space" (e.g., office, gyms), \cite{fan_delque_tvt2011} defines \textit{geo-community} concept and matches each query with a particular geo-community.
Hence, the first question boils down to selecting relays with high probability of visiting the target geo-community.
As \cite{fan_delque_tvt2011} aims to keep the search cost minimal, 
searching node employs two-hop routing and determines the relays, which thereby reduces the issue of when to stop the search.
Deciding when to stop search is nontrivial as the search follows several paths and whether the searching node has already discovered a response is not known by the relaying nodes. Hyyti{\"a} \textit{et al.}~\cite{hyytia2014searching} model the expected search utility with increasing hop count and then finds the optimal hop, while \cite{pitkanen2009searching} estimates the number of nodes having received a query and possible responses by using the node degrees. 
In our work, we showed that simple schemes via information sharing can stop search timely and maintain similar performance to that of an oracle due to the small world nature of the studied networks.

Different from all these above works, main focus of our paper is more on a fundamental question: 
\textit{how does the hop limitation affect the search performance?} 
While this question has been  
explored in general networking context, 
content-centricity and the time constraints 
require a better understanding of flooding in the context of search. 
Therefore, we first provided insights on search on a simplified setting and next
analyzed the effect of various parameters, e.g., time and real mobility traces, via extensive simulations.

In the literature, several works focus on two-hop forwarding in which the source node replicates the message to any relay and the relays can deliver the message only to the final destination~\cite{Grossglauser_TON2002,Chaintreau_TOM2007, al2008performance}. 
Grossglauser \textit{et al.} in their seminal work~\cite{Grossglauser_TON2002} show that two hops are sufficient to achieve the maximum throughput capacity in an ideal network with nodes moving randomly. 
The capacity increase is facilitated by the reduced interference on the links from source to relay and relay to the destination. 
Chaintreau \textit{et al.}~\cite{Chaintreau_TOM2007} assess this two-hop forwarding scheme in a DTN scenario with power-law inter-contact times and 
employ an \textit{oblivious forwarding algorithm} (e.g., memoryless routers that do not use any context information such as contact history). 
Similarly, \cite{boldrini2012less} focuses on a DTN with power-law distributed inter-contact times and derives the conditions (i.e. range of Pareto shape parameter) under which message delivery has a finite delay bound for both two-hop and multi-hop oblivious algorithms. 
The authors show that 
``\textit{as long as the convergence of message delivery delay is the only concern, paths longer than two-hops do not help convergence}'' as two hops are sufficient to explore the relaying diversity of the network~\cite{boldrini2012less}. 
Another work supporting two-hop schemes is \cite{fan_delque_tvt2011} which shows that two-hop search is favourable for opportunistic networks --a resource-scarce setting,  
as ``\textit{one-hop neighbors are able to cover the most of the network in a reasonable time}" in a network with \textit{sufficiently many} mobile nodes.
In our work, we include those cases when longer paths still yield (some) performance improvement. 
Finally, our work supports the conclusion of \cite{Chaintreau_DiameterCoNext2007}, which theoretically proves the \textit{small-world} in human mobile networks.

Our work is closely related to the $k$-hop flooding 
\cite{vojnovic2011hop} which models the spread of flooded messages in a random graph. 
Unlike \cite{vojnovic2011hop}, we focus on content search in a realistic setting, and using real mobility traces~(both a human contact network and a vehicular network) we provide insights on the effect of increasing hop count on the search success, delay, and cost under various content availability and tolerated waiting time settings. 
Despite the differences, it is worthwhile noting that our results agree with the basic conclusions of \cite{vojnovic2011hop}.

\section{Conclusions}\label{sec:Conclusions}
Given the volume of the content created, downloaded, and stored in the mobile devices, 
efficient opportunistic search is paramount to make the remote content accessible to a mobile user.
Current schemes mostly rely on routing based on hop limitations. 
To provide insights about the basics of such a generic search scheme, we focused on a hop-limited search in mobile
opportunistic networks. First, by modelling only the forward path, we
showed that 
(i) the second hop and following few hops bring the
highest gains in terms of forward path success ratio %
and
(ii) compared to single-hop delivery, increase in hop count leads to shorter search time and after
a few hops search time tends to stabilize. 
Next, we revisited these findings via
simulations of the entire search process. While simulations validated
our claim for the forward path, we observed that return path on average requires
longer time and more hops. %
Moreover, our results do not indicate strong correlation between the return and
forward paths. Finally, we showed that search completes in less than
five hops in most cases. This is attributed to the small diameter of
the human contact network which has also a positive impact on search
cost; nodes are informed about search state quickly and stop
propagation of obsolete messages.
 Our simulations validated that increasing hop count to several hops accelerates the search and later search completion time stabilizes.

 As future work, we will design a search scheme that adapts the hop count of query and response paths based on the observed content availability and popularity. 
 Moreover, we believe that content-centric approaches should be paid more attention to implement efficient search schemes in mobile opportunistic networks.

\section*{Acknowledgements}
This work was supported by the Academy of Finland in the PDP project (grant no.\ 260014).
\bibliographystyle{plain}

\appendix
\section{Derivation of Search Success}\label{appendix}
Given that the message is received by $K$ nodes and each node has probability $\avail$ of holding the requested content, the probability that an initiated query reaches one of the content provider(s) equals to the probability that at least one of the $K$ nodes has the content. 
We denote then the forward success ratio as:
\begin{align}
\textrm{Forward success ratio} = 1-(1-\alpha)^K \label{eq:Pfw_appendix}.
\end{align}
If we consider the whole search path with the assumption that $K$ nodes at maximum holds the query and only $K^{'}$ copies are allowed for the response message, we calculate the search success ratio, i.e., both steps are completed, as:
\begin{align}
 P_{s}&= \sum_{k=1}^{K} Pr\{\textrm{k content providers are discovered}\} \nonumber\\
&\times Pr\{\textrm{at least one of k responses reaches } n_s\} \label{eq:Ps_appendix}.
\end{align}

The first factor in (\ref{eq:Ps_appendix}), corresponding to the event that $k$ content providers are discovered, obeys Binomial distribution with parameters $(K,\avail)$.
The second factor is conditioned on $k$, the number of content providers reached by the query replicas. 
Out of these $k$ responses, we calculate the probability that at least one of them reaches the destination, i.e., the searching node. 
With the assumption that $K^{'}$ copies of the response is allowed on the return path, to calculate the probability of finding $n_s$, we find the probability that none of the $k$ responses reaches $n_s$.
Since there are $(\Nodes-1)$ nodes a response message created by a particular content provider can reach and $K^{'}$ selection without replication, the probability that a response reaches $n_s$ equals to: $$\gamma=\frac{K'}{N-1}.$$
Substituting these formulations into (\ref{eq:Ps_appendix}), we find:
\begin{align}
P_{s}&{=}\sum_{k=1}^{K} \binom{K}{k}\avail^k (1{-}\avail)^{K-k} \left(1{-}(1-\gamma)^k\right)\label{eq:app_0}. 
\end{align}
Separating (\ref{eq:app_0}) into two parts, we find:
\begin{align}
P_{s}{=}&\sum_{k=1}^{K} \binom{K}{k}\avail^k (1{-}\avail)^{K-k} 
		-&\sum_{k=1}^{K} \binom{K}{k}\avail^k(1{-}\avail)^{K-k} (1-\gamma)^k. \label{eq:app_1} 
\end{align}

Next, we notice that each summation term above can be compactly written by the help of the binomial theorem, %
 which is: $$(x+y)^n = \sum_{k=0}^{n}x^ky^{n-k}.$$ 
Taking into account that the summation in (\ref{eq:app_1} ) starts from 1, we simplify the first term in (\ref{eq:app_1}) as follows:
\begin{align}
\sum_{k=1}^{K} \binom{K}{k}\avail^k (1{-}\avail)^{K-k} &= (\avail+1-\avail)^K - \binom{K}{0}\avail^0 (1{-}\avail)^{K-0} \nonumber\\ %
&= 1-(1{-}\avail)^{K}.  \label{eq:app_first_term}
\end{align}
Applying the same expansion for the second term in (\ref{eq:app_1}), we find:
\begin{align}
&\sum_{k=1}^{K} \binom{K}{k}\avail^k(1{-}\avail)^{K-k} (1-\gamma)^k = \sum_{k=1}^{K} \binom{K}{k}\left(\avail-\avail\gamma\right)^k (1{-}\avail)^{K-k} \nonumber \\
&= (\avail-\avail\gamma+1{-}\avail)^K - \binom{K}{0}\left(\avail-\avail\gamma\right)^0(1{-}\avail)^{K-0} \nonumber\\
&= (1-\avail\gamma)^K - (1{-}\avail)^{K}.  \label{eq:app_second_term}
\end{align}
Substituting (\ref{eq:app_first_term}) and (\ref{eq:app_second_term}) into  (\ref{eq:app_1}), we find:
\begin{align}
P_{s}&= 1-(1{-}\avail)^{K} - \left((1-\avail\gamma)^K - (1{-}\avail)^{K}\right) \nonumber,
\end{align}
which gives us the search success probability:
\begin{align}
P_{s}= 1- (1-\avail\gamma)^{K} \label{eq:appendix_Ps_final}.
\end{align}
\end{document}